\documentclass[prb,reprint,aps,twocolumn,nopacs,superscriptaddress]{revtex4-2}
\usepackage[utf8]{inputenc}
\usepackage[american,]{babel}
\usepackage[T1]{fontenc}
\usepackage[pdftex]{graphicx}  
\usepackage{graphicx, xcolor}
\usepackage{dcolumn}
\usepackage{bm}
\usepackage{amsmath,amsthm,amssymb}
\usepackage{hyperref}
\usepackage[T1,T2A]{fontenc}
\usepackage{xcolor}
\hypersetup{colorlinks,bookmarksopen,bookmarksnumbered,
    citecolor=blue,
    linkcolor=blue,
    pdfstartview=false,
    urlcolor=blue}
\usepackage{graphicx}
\usepackage{braket}
\usepackage{wrapfig}
\usepackage{soul}
\usepackage{mathtools}
\usepackage{float}
\usepackage{tabularx}

\renewcommand{\vec}{\mathbf}

\begin{document}
\title{Measurement-induced phase transitions in $(d+1)$-dimensional stabilizer circuits}

\author{Piotr Sierant}
\affiliation{ICFO-Institut de Ci\`encies Fot`oniques, The Barcelona Institute of Science and Technology, Av. Carl Friedrich Gauss 3, 08860 Castelldefels (Barcelona), Spain}
\author{Marco Schir\`o}
\affiliation{JEIP, USR 3573 CNRS, Coll\`{e}ge de France, PSL Research University, 11 Place Marcelin Berthelot, 75321 Paris Cedex 05, France}
\author{Maciej Lewenstein}
\affiliation{ICFO-Institut de Ci\`encies Fot`oniques, The Barcelona Institute of Science and Technology, Av. Carl Friedrich Gauss 3, 08860 Castelldefels (Barcelona), Spain}
\affiliation{ICREA, Passeig Lluis Companys 23, 08010 Barcelona, Spain}
\author{Xhek Turkeshi}
\affiliation{JEIP, USR 3573 CNRS, Coll\`{e}ge de France, PSL Research University, 11 Place Marcelin Berthelot, 75321 Paris Cedex 05, France}
\date{\today}
\begin{abstract}
The interplay between unitary dynamics and local quantum measurements results in unconventional non-unitary dynamical phases and transitions. In this paper we investigate the dynamics of $(d+1)$-dimensional hybrid stabilizer circuits, for $d=1,2,3$. We characterize the measurement-induced phases and their transitions using large-scale numerical simulations focusing on entanglement measures, purification dynamics, and wave-function structure. Our findings demonstrate the measurement-induced transition in $(d+1)$ spatiotemporal dimensions is conformal and close to the percolation transition in $(d+1)$ spatial dimensions. 
\end{abstract}

\maketitle

\section{Introduction}
\label{sec:intro}
Understanding the propagation of quantum information in a many-body quantum system coupled to an environment is a central problem at the interface between the quantum foundation, statistical mechanics, high energy physics, and condensed matter theory~\cite{breuer2002theory,nielsen00}.
In the most straightforward formulation, the environment is a measurement apparatus that continuously probes the system's local degrees of freedom -- a setup that plays a pivotal role in comprehending noisy intermediate-scale quantum devices~\cite{Preskill_2018} and, in general, for quantum simulation \cite{Fraxanet22}.
The dynamics combine quantum unitary evolution and measurements in such settings, resulting in stochastic quantum trajectories~\cite{carmichael2009open,Dalibard92,Molmer93, gardiner2004quantum,wiseman2009quantum,daley2014quantum,basche1995direct,gleyzes2007quantum,vijay2011observation,robledo2011highfidelity,minev2019to}.
The unitary evolution introduces coherences and scrambles quantum information throughout the system while local quantum measurements disentangle and localize the degrees of freedom according to the Born-Von Neumann postulate.
For a many-body system, this competition results in measurement-induced phase transitions (MIPTs) for the states along quantum trajectories, visible in non-linear functionals of the state~\cite{skinner2019measurementinduced,li2018quantum,li2019measurementdriven,chan2019unitaryprojective,buchhold2021effective,muller2022measurementinduced,minoguchi2022continuous,turk3,ziocan,yang2022entanglement,trebst,Kalsi_2022,kells2021topological,fleckenstein2022nonhermitian,gullans,boorman2022diagnostics,turk2,legal,Biella2021manybodyquantumzeno,szyniszewski2020universality,barratt2022transitions,zabalo2022infinite,barratt2022field,lunt2020measurement,zabalo2022operator,Zhang2022universal,han2022measurement,iaconis2021multifractality,jian2021meas,noncrit,turk,Sierant2022dissipativefloquet,tony,moessner,pizzi}.

Random unitary circuits are minimal models for investigating the dynamical phases of monitored quantum systems~\cite{fisher2022quantum,lunt2021quantum,potter2021entanglement}. 
In such circuits, the unitary part of the dynamics  is built out of random unitary gates drawn from a specific distribution --- hence lacking the structural constraints present in realistic Hamiltonians~\cite{nahum2017quantum,nahum2018operator,zhou2019emergent,khemani2018operator,chan2018solution}. The non-unitary dynamics stem from local projective (or generalized) measurements that act stochastically at every space-time point with a specified rate.
In (1+1)-dimensional systems, intensive analytical and numerical investigations for Haar and stabilizer random circuits lead to a well-understood phase diagram~\cite{szyniszewski2019entanglement,sand2021entanglement,shi2020entanglement,sang2021measurementprotected,Lavasani2021,block2022measurementinduced,Noel_2022,ibmmotta,turkeshi2021measurementinduced,jian2020measurementinduced,Bao_2021,lopez2020meanfield,agrawal2021entanglement,ippoliti2021entanglement,gullans2020scalable,Fidkowski2021howdynamicalquantum,sharma2022measurementinduced,lang2020entanglement,klocke2022topological}. 
At a low-measurement rate, quantum correlations are resilient to the disturbing action of local measurements resulting in a quantum error-correcting (QEC) phase  ~\cite{bao2020theory,choi2020quantum}. In contrast, frequent measurements deteriorate the quantum correlations and freeze the dynamics onto a reduced manifold, yielding a quantum Zeno (QZ) phase~\cite{li2018quantum,Georgescu2022}.
A measurement-induced critical point separates these dynamical phases, exhibiting a rich non-unitary conformal field theory (CFT) with geometric features~\cite{skinner2019measurementinduced,zabalo2020critical,vasseur2019entanglement,li2021conformal}.

Different but related aspects of these dynamical phases have been identified depending on the initial condition and the probes used to investigate the transition. 
Entanglement measures~\cite{li2018quantum,li2019measurementdriven,skinner2019measurementinduced} show that the entanglement entropy of a pure state in the QEC phase follows a volume law akin to ergodic systems~\cite{Alessio16}.  In contrast, in the QZ phase, the entanglement entropy obeys an area law similar to ground states ~\cite{Eisert10} or eigenstates in many-body localized systems~\cite{abanin2019colloquium}. 
In contrast, when the system starts evolving from a mixed state, the environment monitoring reduces the entropy of the system density matrix.
This allows identifying the QEC phase with a mixed phase with exponentially divergent purification time, distinct from a pure phase with polynomial purification time, corresponding to the QZ phase. From that point of view, the measurement-induced phase transition becomes a purification transition~\cite{gullans2020dynamical}.
Lastly, Ref.~\cite{sierant2022universal} has characterized the measurement-induced criticality by directly inspecting the structure of wave functions on individual quantum trajectories. In the QEC phase, the participation entropy develops a non-zero sub-leading term determined by details of the unitary part of dynamics, which, in contrast, vanishes in the QZ phase. 

Beyond one spatial dimension, quantum correlations and information propagation are less comprehended. 
For random Haar circuits that contain generic local unitary gates, numerical methods are prohibitive due to the exponential (classical) resources needed for their simulation. Ref.~\cite{skinner2019measurementinduced} proposed a phenomenological geometric minimal-cut picture for entanglement spreading in this generic case in the presence of random measurements (see also Ref.~\cite{nahum2017quantum}). 
Conversely, the unitary gates in stabilizer circuits are chosen from the Clifford group which is a discrete sub-group of the full unitary group, allowing for simulation of such circuits with classical polynomial resources even in the presence of projective measurements~\cite{Gottesman98,aaronson2004improved}. This enables numerical investigation of the transition between QEC and QZ phases in higher dimensions. In particular, Ref.~\cite{turkeshi2020measurementinduced,lunt2021measurementinduced,lavasani2021topological} studied the (2+1)-dimensional systems but found conflicting results, likely due to the limited system sizes in their computations or the choice of observable used to locate the transition.
Therefore, it is desirable to employ large-scale numerical simulations and clarify the phase diagram in higher dimensions for stabilizer circuits.

This paper revisits the MIPT in $(d+1)$-dimensional stabilizer circuits.
We utilize large-scale numerics up to $N=32768$ qubits to unveil a precise characterization of the critical exponents of the MIPT for $d=1,2$, as well as consider the previously unexplored case $d=3$. We provide a coherent picture of measurement-induced criticality by probing the three aspects of dynamics of the circuits that change between QEC and QZ phases: the entanglement of the pure state, the purification of initially maximally mixed state as well as the structure of the wave function along quantum trajectories.  We test the robustness of our results by considering different observables and various spatiotemporal architectures of the circuits. Furthermore, we investigate the bulk-boundary critical properties by considering ancilla qubits coupled to our circuits.
In the spirit of Ref.~\cite{li2021statistical_1}, we suggest that the resulting statistical field theory at the transition between QEC and QZ phases is that of a perturbed percolation model, and hence the critical point is conformal.
This ansatz is compatible with our numerical findings. We show that the critical exponents for (1+1)D circuits are close but distinct from the percolation transition in $2$ dimensions, whereas the results for $(d+1)$-dimensional circuits for $d=2,3$ are compatible with the percolation transition in $(d+1)$ spatial dimensions,
see Table.~\ref{tab:critical_exponents} for the summary of the results.

The rest of this work is structured as follows. First, in Sec.~\ref{sec:overview} we give a brief overview of our approach and summarize the obtained critical exponents for MIPT in $(d+1)$-dimensional circuits.
Sec.~\ref{sec:methods} describes the architectures of stabilizer circuits that are employed in our investigations of MIPTs. In Sec.~\ref{sec:results} we detail the entanglement entropy measures and extract the universal properties of the MIPT through finite-size scaling.
Sec.~\ref{sec:purtrans} contains results about the purification aspect of MIPT while
Sec.~\ref{sec:struct} describes the wave-function structure change at the transition between QEC and QZ phases. Sec.~\ref{sec:blkbnd} is devoted to the investigation of the bulk-boundary features of the MIPT through coupling with ancillary qubits.
Finally, Sec.~\ref{sec:results} discusses our results and illustrates the conformal character of the critical point, while Sec.~\ref{sec:conclusion} concludes the manuscript with some outlooks. In the Appendices, we detail the most technical aspects of our work.

\section{Overview of the results}
\label{sec:overview}
Let us first present the rationale of our analysis and summarize our main findings. After motivating the choice of observables for the quantum trajectories, we briefly discuss how they describe the measurement-induced phases and transitions, highlighting the key results. 
In summary, we establish that $(d+1)$ spacetime dimensional stabilizer circuits have a MIPT with universal behavior close to a percolation field theory in $(d+1)$ spatial dimensions. In particular, for $d=1$ the stabilizer circuit critical exponents are distinct and within five error bars from the 2D percolation exact results, while for $d=2$ ($d=3$) \emph{all} our estimates are compatible within one error bar with those of 3D (4D) percolation~\ref{tab:critical_exponents}. 
In addition, we report evidence suggesting an emergence of a conformal field theory right at the MIPT in $(d+1)$-dimensional stabilizer circuits for $d=2$ and $d=3$.

\paragraph{Observables of interest and quantum trajectories} This manuscript considers quantum circuits build from: (i) random unitary gates (drawn uniformly from the Clifford group), (ii) projective measurements onto the local qubit basis. 
In particular, a circuit realization $K_\mathbf{m}$ (with implicit depth $t$, also referred to as time) stems from the choice of Clifford gates, the spacetime location of measurements, and the measurement outcomes. 
Starting from the initial state $\rho_0$, the final state is given by
\begin{equation}
    \rho_\mathbf{m} = \frac{K_\mathbf{m} \rho_0 K^\dagger_\mathbf{m}}{\mathrm{tr}(K^\dagger_\mathbf{m} K_\mathbf{m}\rho_0)}.\label{eq:deftraj}
\end{equation}
We denote the collective average by $\mathbb{E}_\mathbf{m}$. 
The trajectory nature of Eq.~\eqref{eq:deftraj} translates to statistical properties of an observable $\Xi$ on the state $\rho_\mathbf{m}$. If $\Xi$ is linear in the density matrix, it is easy to see that 
\begin{equation}
    \mathbb{E}_\mathbf{m}[ \Xi[\rho_\mathbf{m}] ] = \Xi[\mathbb{E}_\mathbf{m}(\rho_\mathbf{m})].
\end{equation}
Notably, the above equation does not hold when $\Xi$ is a non-linear function of the density matrix.
Consider for instance the purity $\Xi[\rho]=\mathrm{tr}(\rho^2)$, and a pure state quantum trajectory. Since the average state $\overline{\rho}=\mathbb{E}_\mathbf{m}(\rho_\mathbf{m})$ is mixed, we have $1=\mathbb{E}_\mathbf{m}[ \Xi[\rho_\mathbf{m}]  ]\neq \Xi[\overline{\rho}]\le 1$. 
In particular, the average state evolves through a quantum channel and is insensible of the trajectory registry $\mathbf{m}$. In other words, the statistical structure of quantum trajectories contains more information (including measurement-induced criticality) than the average state.
We consider entropic observables: being non-linear in the density matrix, they allow us to investigate the various aspects of the MIPT. Crucially, as discussed in the remaining, they are efficiently computable for stabilizer states, allowing extensive numerical characterization of MIPTs in $(d+1)$-dimensional stabilizer circuits. 
In Sec.~\ref{sec:methods} we detail the properties of stabilizer circuits and specify the circuital architectures treated in this manuscript.

\begin{figure}[t!]
    \centering
    \includegraphics[width=\columnwidth]{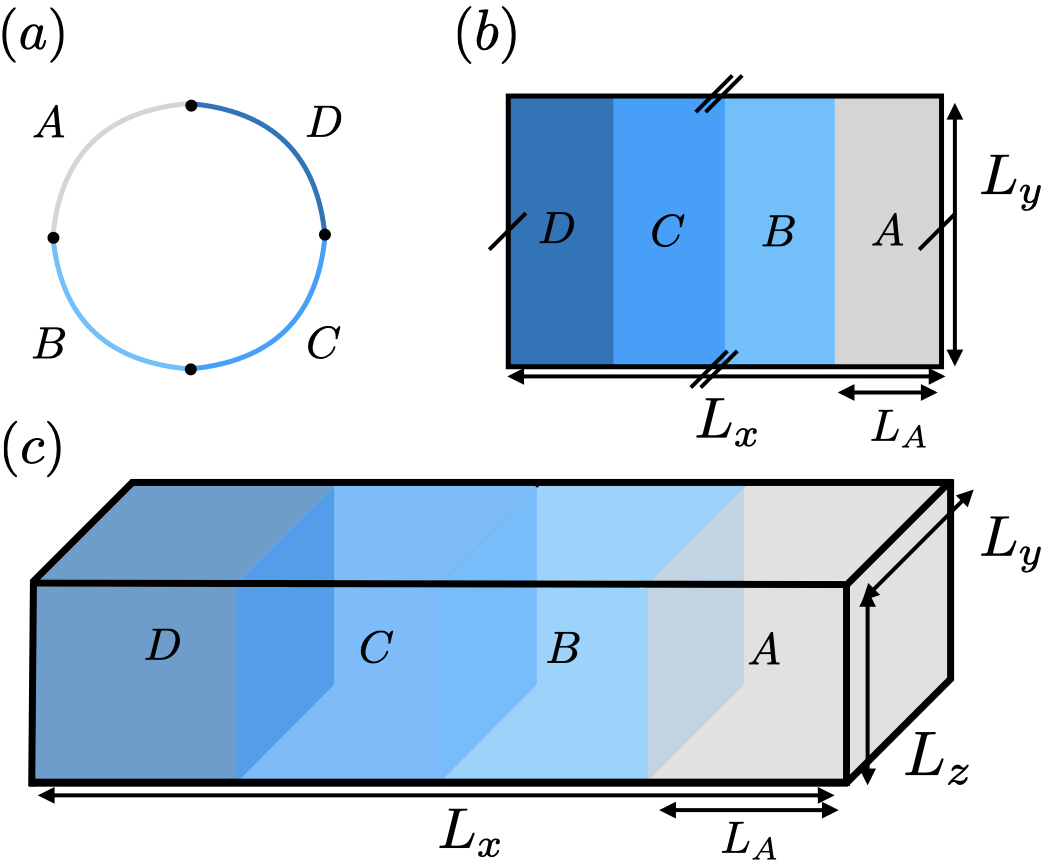}
    \caption{Cartoon of the partitions considered for the entanglement entropy and for the tripartite quantum mutual information (TQMI) for (a) (1+1)-dimensional, (b) (2+1)-dimensional, and (c) (3+1)-dimensional systems (we illustrate only the $d$ spatial dimensions of the circuits). If not specified, we infer periodic boundary conditions in every direction. For the analysis in Sec.~\ref{sec:results} we consider $L_A=L_B=L_C=L_D=L_x/4$, and $L_x=L_y=L_z$. Instead, in Sec.~\ref{sec:dis} we consider $L_y=L_z$ while varying $L_x$ and $L_A$. }
    \label{fig:partition}
\end{figure}

\begin{table*}[t!]
    \centering
    \begin{tabular}{cccccccccc}
        \hline \hline
        & \multicolumn{3}{c}{Stabilizer circuits} &  &  \multicolumn{3}{c}{Classical percolation} \\
        \cline{2-4} \cline{6-8}
        {\raisebox{1.5ex}[0pt]{ Exponent }} & {1+1D} & {2+1D} & {3+1D} & &  {2D} & {3D} & {4D}\\
        \hline
        $\nu$ & $\quad$ 1.265(15) $ \quad$ & $\quad$ 0.87(2) $\quad$ & $\quad$  0.68(2) $\quad$  & $\quad $ & $\quad$  1.333 $\quad$ &  $\quad$ 0.8774(13) $\quad$  & $\quad$ 0.686(2)  $\quad$  \\
        $\eta$ & 0.212(4) & -0.02(3) & -0.04(5) & & $ 0.208$  & $- 0.03(1)$ & -0.084(4) \\
       $\eta_{\parallel}$ & 0.70(2) & 0.99(4) & 1.5(2)  &  &  $ 0.667$ & 1.08(10) &  1.37(13) \\
        $\eta_{\perp}$ & 0.461(8) & 0.43(5) & 0.42(10) &  &  $  0.438$ & 0.5(1) & 0.65(7)\\
        $\beta$ & 0.129(8)
        & 0.44(1) & 0.60(3) &  &   $ 0.139$ & 0.429(4) & 0.658(1)\\
        $\beta_s$ & 0.46(2)
        & 0.86(2) & 1.14(9) & &   $ 0.444$ & 0.854(2) & 1.09(8)\\
        $z$ & 1.00(1) & 1.01(2) & 1.02(4) &  & 1 & 1 & 1\\
        \hline \hline
    \end{tabular}
    \caption{ Summary of the critical exponents for the stabilizer hybrid quantum circuits and comparison with classical percolation theory. Systems in $(d+1)$ space-time dimensions should be compared with $(d+1)$ spatial dimensions classical percolation theory. The critical exponents of 2D percolation are exact and follow by conformal field theory~\cite{DiFrancesco:639405}, and here we report the truncation of their rational values to facilitate the comparison. The value $z=1$ arises from the conformal invariance of the percolation field theory. 
    For the 3D and 4D percolation critical exponents, we refer to  Ref.~\cite{phifiveloops,Koza_2016,BenAlZinati2018} for the bulk properties, and for Ref.~\cite{surface1,surface2,surface3,phifive} for the surface properties.~\label{tab:critical_exponents} }
\end{table*}

\paragraph{Entanglement transition.}
For a given density matrix $\rho$, the von Neumann entropy is defined as
\begin{equation}
    S(\rho) \equiv - \mathrm{tr} (\rho\log_2 \rho)
    \label{eq:entdef0}
\end{equation}
where the trace is over the available degrees of freedom. 
The entanglement entropy is defined for a pure state $\rho=|\Psi\rangle\langle \Psi|$ and a bipartition $A\cup A_c$ of the system as
\begin{equation}
    S_A \equiv S(\rho_A),\qquad \rho_A = \mathrm{tr}_{A_c} \rho, 
    \label{eq:entdef}
\end{equation}
where $\mathrm{tr}_{A_c}$ is the trace over degrees of freedom of subsystem $A_c$.
Within these conditions, the entanglement entropy is a measure of distillable quantum information~\cite{nielsen00}.

We consider a qubit system on a $d$-dimensional hyper-parallelepiped $\Lambda$  and a subsystem $A$ is extensive ($|A|\sim |\Lambda|$, where $|X|$ is the total number of sites in $X$. See Fig.~\ref{fig:partition}, where $A_c= B\cup C \cup D$). 
Due to the trajectory nature of the system, in the following, we refer to the entanglement entropy as its average value $S_A= \mathbb{E}_\mathbf{m}[S(\rho_{\mathbf{m},A})]$, with $\rho_{\mathbf{m},X}=\mathrm{tr}_{X_c} (\rho_\mathbf{m}$).

Then, the MIPT manifests itself by an abrupt change in the scaling of the entanglement entropy as the measurement rate $p$ is tuned across the critical measurement rate $p_c$~\cite{li2018quantum,li2019measurementdriven,skinner2019measurementinduced, turkeshi2020measurementinduced,lunt2021measurementinduced,lavasani2021topological}. For $p<p_c$, the system is in the QEC phase, and the quantum information encoded in the system is resilient to the local measurements, yielding a volume-law stationary entanglement entropy $S_A\propto |A|$. Conversely, in the QZ phase ($p>p_c$), the measurements deteriorate quantum correlations restricting the creation of quantum entanglement.
As a result, the entanglement entropy follows an area-law~\cite{Eisert10} $S_A\propto |\partial A|$, with $\partial X$ being the boundary of $X$. In Sec.~\ref{sec:results}, we indeed observe numerically the volume-law and area-law scaling regimes for the entanglement entropy $S_A$ in stabilizer circuits in $(d+1)$-dimensions for $d=1,2,3$.

The entanglement entropy at the transition $p=p_c$ typically presents non-trivial system size dependence (for instance, a logarithmic correction for $d=1$). A measure better suited for the analysis of the entanglement transition is the tripartite quantum mutual information (TQMI)~\cite{zabalo2020critical} defined as
\begin{equation}
    I_3 = S_A + S_B + S_C - S_{A\cup B} - S_{A\cup C} - S_{B\cup C} + S_{A\cup B\cup C}\;.\label{eq:i3}
\end{equation}
Here $\{A,B,C,D\}$ is the quadripartition of $|\Lambda|$ in Fig.~\ref{fig:partition}, where $A,B,C,D$ are neighboring regions of equal size. 
Then, the scaling of the entanglement entropy $S_A$ in QEC and QZ phases implies the following scaling for the TQMI
\begin{equation}
    I_3(p,L) = \begin{cases}
    O(L^{d}),& p<p_c,\\
    O(1), & p=p_c,\\ 
    \sim e^{-\gamma(p) L}, & p>p_c,
    \end{cases}
\end{equation}
where $\gamma(p) >0$ is a $p$ dependent constant. In particular, in the scaling limit, the curves $I_3(p)$ for different system sizes $L$ cross at criticality, allowing us to pinpoint the transition point. 
To analyze the properties of the transition, we assume an emergence of a power-law divergent  length scale at the transition
\begin{equation}
    \xi \sim \frac{1}{|p-p_c|^{\nu}},
\end{equation}
where $p_c$ is the critical measurement rate, and $\nu$ is a critical exponent. This leads us to the scaling ansatz around the transition
\begin{equation}
    I_3(p,L)\simeq F[(p-p_c)L^{1/\nu}],
    \label{eq:I3ansatz}
\end{equation}
where $F(x)$ is a system size independent function. 

As we detail in Sec.~\ref{sec:results}, Eq.~\eqref{eq:I3ansatz} allows us to obtain excellent data collapses for (1+1)D, (2+1)D and (3+1)D stabilizer circuits, with the critical exponents $\nu$ summarized in Table~\ref{tab:critical_exponents}~\footnote{We have performed similar data collapses for the entanglement entropy. While finding compatible results, the data collapse estimates have larger error bars than $I_3$ due to the more stringent finite-size effects. We decided not to present these further numerical verifications since they do not add novel insights.}.
While we postpone a more detailed discussion of those results to Sec.~\ref{sec:dis}, we note  that the critical exponent $\nu$ for (1+1)D is close but different from 2D classical percolation, whereas the exponents $\nu$ describing entanglement transition in (2+1)D and (3+1)D circuits are compatible with the exponents of 3D and 4D classical percolation.

\paragraph{Purification transition.} Another facet of measurement-induced criticality reveals itself when one initializes the system in a mixed state  $\rho$ \cite{gullans2020dynamical}. In such a case, discussed in Sec.~\ref{sec:purtrans}, the measurements will increasingly resolve the state and lead to complete purification~\cite{Fidkowski2021howdynamicalquantum}. However, the timescale at which the purification happens varies considerably between the QEC and the QZ phases. In the former, the coherences obstruct the environment from the complete resolution of the information content, and the state purifies at times of the order the Hilbert space dimension (meaning that the environment has to probe virtually all available degrees of freedom). In the latter case, the measurements resolve information in a polynomial timescale in system size in the QZ phase~\footnote{The above arguments justify the alternative naming for the QEC and the QZ phase, as the mixed and the pure phase, respectively.}. 

Starting from a fully mixed state $\rho_0\propto \openone$, we calculate the von Neumann entropy $S(\rho_{\mathbf{m}})$ of the time-evolved density matrix (cf. Eq.~\eqref{eq:entdef0}), and average the results obtaining the average entropy $S_\mathrm{pur}=\mathbb{E}_\mathbf{m}[S(\rho_{\mathbf{m}})]$. In Sec.~\ref{sec:purtrans}, we observe that at the MIPT, $p=p_c$, the entropy $S_\mathrm{pur}$ becomes a universal function of $t/L^z$ where $t$ is the circuit depth and $z$ is a dynamical critical exponent and find that, within estimated error bars, $z=1$ for (1+1)D, (2+1)D and (3+1)D stabilizer circuits, as shown in Table~\ref{tab:critical_exponents}. This value of the dynamical critical exponent stems from scaling invariance, hence it is compatible with a conformal invariance at the critical point. Moreover, we demonstrate that data for the average entropy
$S_\mathrm{pur}$ at circuit depth $\tau = \alpha L$ (with $\alpha$ is a constant) 
can be collapsed using the ansatz 
\begin{equation}
    S_\mathrm{pur} = G((p-p_c)L^{1/\nu}, t/L),
    \label{eq:pur}
\end{equation}
with the same values of the critical exponents $p_c$, $\nu$ as the ones obtained in the analysis of the tripartite mutual information (cf. Eq.~\eqref{eq:I3ansatz}), and with the dynamical critical exponent assumed to be $z=1$, see Sec.~\ref{sec:purtrans}. 
The above conclusions, and the coincidence of purification and entanglement transition, hold for all the spacetime dimensions considered.
This phenomenon is non-trivial as a class of circuits exhibits qualitatively different behavior between the mixed purification dynamics and the pure entanglement dynamics~\cite{lu2021spacetime,ippoliti2022fractal,ippoliti2021postselectionfree,claeys2022exact}.

\paragraph{Wave-function structure across MIPT.}
We study the structural properties of stabilizer circuits employing participation entropy. This quantity is a measure of localization of the system wave function in a given basis of many-body Hilbert space and captures changes in wave-function structure across quantum phase transitions~\cite{DeLuca2013,Luitz2014,laflo,luitz,luitz2,luitz3}, as well as in non-equilibrium settings such as many-body localization \cite{Mace19, Luitz20, Pietracaprina21, Solorzano21}. The participation entropy was also recently shown to successfully capture the universal behavior of MIPTs in (1+1)D hybrid quantum circuits~\cite{sierant2022universal}.

Given a pure state $\rho=|\Psi\rangle\langle \Psi|$ the participation entropy is defined as 
\begin{align}
	S_\mathrm{part}(\rho) = - \sum_{\vec{\sigma}}  	\braket{\vec{\sigma} |\rho | \vec{\sigma}} \log_2 	\braket{\vec{\sigma} |\rho | \vec{\sigma}},
	\label{sqdef}
\end{align}
where the sum extends over the full basis $\ket{\sigma}$ of $2^{|\Lambda|}$-dimensional Hilbert space. 
We consider the wave function at individual quantum trajectories and consider the average value $S_\mathrm{part} = \mathbb{E}_\mathbf{m}[S_\mathrm{part}(\rho_\mathbf{m})]$.
It exhibits the scaling 
\begin{equation}
    S_\mathrm{part} = D |\Lambda| + c,
    \label{eq:partdef}
\end{equation}
where $D$ and $c$ are the fractal dimension and the fractal sub-leading term, respectively. The coefficients $D$ and $c$ can be used to analyze further the transition between QEC and QZ phases. In particular, for (1+1)D systems, the fractal subleading term $c$ encodes the universal content of the measurement-induced phase transition~\cite{sierant2022universal,sierant2022universality}, while $D$ is non-universal and phase dependent. 
In Sec.~\ref{sec:struct} we show that the same conclusion holds for (2+1)D and (3+1)D systems and perform a finite size scaling  analysis of the fractal sub-leading term with the ansatz
\begin{equation}
    c = Y((p-p_c)L^{1/\nu}),
    \label{eq:partcol}
\end{equation}
to extract the critical measurement strength $p_c$ and the correlation length exponent $\nu$.
Our estimates of the exponent $\nu$ and the value of $p_c$ are consistent with the results obtained for the entanglement and purification transitions (cf. Table~\ref{tab:critical_exponents}).

\paragraph{Bulk-boundary properties at MIPT.}
To further characterize measurement-induced criticality in $(d+1)$-dimensional stabilizer circuits we obtain, in Sec.~\ref{sec:blkbnd}, the bulk and boundary critical exponents by investigating the purification of ancilla qubits entangled with the system at a certain moment of time. Notably, these local order parameters are experimentally accessible through a few site full tomography~\cite{Noel_2022,ibmmotta}.

We consider the combined framework $|\Phi\rangle \equiv |\Psi\rangle\otimes |a\rangle$, where $|\Psi\rangle$ is the system pure state and $|a\rangle$ is a reference state of an ancilla qubit. The system is entangled with the ancilla at time $t_0$ through a certain unitary operation  and
the combined setup evolves under the quantum circuit which acts non-trivially only on the system qubits.
To characterize the entanglement between the ancilla qubit and the system, we compute the average entanglement entropy of the ancilla qubit $S_\mathrm{anc}$.  The ancilla qubit gets quickly disentangled with the system in the QZ phase, yielding a vanishing value for $S_\mathrm{anc}$ at sufficiently large circuit depths. In contrast, in QEC phase, for $p<p_c$, we find that $S_\mathrm{anc} \propto |p_c-p|^{\tilde\beta}$. Depending on the choice of the initial state of the system, the behavior of $S_\mathrm{anc}$ determines the exponents $\tilde\beta \equiv \beta_s$, i.e. the boundary order parameter, and $\tilde\beta \equiv \beta$, i.e. the bulk order parameter (as we discuss in details in Sec.~\ref{sec:blkbnd}). Table~\ref{tab:critical_exponents} presents our results for the bulk and boundary order parameters $\beta$ and $\beta_s$ for (1+1)D, (2+1)D and (3+1)D stabilizer circuits. Besides the small deviations for (1+1)D case, the exponents $\beta, \beta_s$ for the $(d+1)$-dimensional stabilizer circuits are compatible with percolation theory in $d+1$ spatial dimensions.

Furthermore, we set the measurement rate to be equal to its critical value, $p=p_c$, and use $z=1$ estimated from the entanglement and purification observables to characterize the bulk-bulk ($\eta$), bulk-boundary ($\eta_\perp$), boundary-boundary ($\eta_\parallel$) critical exponents~\cite{lunt2021measurementinduced,zabalo2020critical}. 
In this case, we consider the combined framework $|\Psi\rangle\otimes |a,b\rangle$ where $|\Psi\rangle$ is the system pure state and $|a,b\rangle$ are two ancilla qubits. We entangle the latter at time $t_0$ with two system qubits distant $r$, and, for $t>t_0$, we study the behavior of the average bipartite quantum mutual information $I_2 = \mathbb{E}_\mathbf{m}[I_2(\rho_{\mathbf{m},a,b})]$ where
\begin{equation}
    I_2(\rho_{\mathbf{m},ab}) \equiv S(\rho_{\mathbf{m},a}) + S(\rho_{\mathbf{m},b}) - S(\rho_{\mathbf{m},ab}).
    \label{eq:i2def}
\end{equation}
Using $I_2$, we access all the anomalous critical exponents  $\tilde \eta =\{ \eta,\eta_\perp,\eta_\parallel \}$ by considering circuits with different boundary conditions and different starting times $t_0$.
The exponents $\tilde \eta$ are determined by the behavior of the bipartite quantum mutual information
\begin{equation}
    I_2(t,r) = \frac{1}{r^{d-1+\tilde\eta}} G( (t-t_0)/r) 
\end{equation}
where $\tilde\eta\in \{\eta,\eta_\perp,\eta_\parallel\}$ depends on the setup, as specified in Sec.~\ref{sec:blkbnd}.
Table~\ref{tab:critical_exponents} presents our results for the various $\tilde\eta$ in (1+1)D, (2+1)D and (3+1)D stabilizer circuits. Besides the small deviations for (1+1)D case, for the $(d+1)$-dimensional stabilizer circuits, these exponents are compatible with percolation theory in $d+1$ spatial dimensions.

\paragraph{Entanglement entropy behavior at the transition.}
Lastly, we study the entanglement entropy at the transition and report evidence of a conformal critical behavior. Specifically, we consider (2+1)D and (3+1)D circuits and calculate the torus entanglement entropy introduced in Ref.~\cite{Witczak17} for (2+1)D and (3+1)D systems with conformal invariance. 

Assuming that a conformal field theory describes the critical point, the scaling of the entanglement entropy at $p=p_c$ for region $A$  in $d>1$ dimensions is given by 
\begin{equation}
    S(A) = \mathcal{B} |\partial A| - \chi + O(1/ |\partial A|),\label{eqtorus} 
\end{equation}
where $\mathcal{B}$ is a non-universal coefficient, and $\chi$ is a universal constant. 
The latter can be analytically computed from an exemplary model with conformal invariance at $d>1$, i.e. the Extensive Mutual Information model~\cite{Casini05,Casini09}. 
Following the guideline of Ref.~\cite{Witczak17}, in Sec.~\ref{sec:dis}, we numerically compute the torus entanglement entropy fixing the aspect ratio $b=L_x/L_y$ ($L_y=L_z$) (see Fig.~\ref{fig:partition}) for the (2+1)D and (3+1)D circuits,  and comparing it with the analytical predictions within the CFT. 
We find that the analytical predictions of CFT match the numerical data, and observe diminishing discrepancies with increasing system size.

\section{Stabilizer circuits}
\label{sec:methods}
Stabilizer states and circuits are pivotal in quantum error correction~\cite{nielsen00,Gottesman98} and in the study of MIPT~\cite{li2018quantum,li2021statistical_1,li2021statistical_2}.
This is due to a combination of their classical simulability via the Gottesman-Knill theorem~\cite{aaronson2004improved}, ability to host an extensive entanglement \cite{Emerson03, Znidaric20} as well as the fact that they form a unitary-2 design \cite{DiVincenzo02} (see also \cite{Webb16,Zhu17}).
This section reviews stabilizer circuits and defines the circuital architectures considered in the manuscript. The system is defined on the $d$-dimensional 
hyper-parallelepiped lattice $\Lambda$, with $\vec{i}\in \Lambda$ its sites. We shall consider both periodic and open boundary conditions, depending on the observables of interest. 

In our numerical calculations, we employ the state-of-the-art package Stim~\cite{Gidney21} which relies on the ideas of~\cite{Gottesman98,aaronson2004improved, Koenig14}.
The package Stim allows for an efficient numerical simulation of stabilizer circuits with time and memory complexity scaling polynomially in system volume $|\Lambda|$. Below, we outline the main ideas underlying simulation of stabilizer circuits and provide details of spatiotemporal architectures of circuits employed in this work.

\subsection{Round-up on stabilizer circuits}
\label{sub:roundup}
Here, for self-consistency and completeness, we briefly review the crucial properties of stabilizer circuits and how they are efficiently simulatable via the Gottesman-Knill theorem~\cite{aaronson2004improved}. 

A \emph{pure} stabilizer state $|\Psi\rangle$ on $N=|\Lambda|$ qubits is characterized by a group of Pauli strings $\mathcal{G}_\Psi=\{g\}$ such that $g|\Psi\rangle = |\Psi\rangle$ and $|\mathcal{G}_\Psi|=2^N$. In this paper, we shall consider $\mathcal{G}_\Psi$ to be a subgroup of all the Pauli strings acting on the lattice $\Lambda$, whose elements are in the form
\begin{equation}
\label{eq:defpauli}
    g = e^{i\pi \phi} \prod_{\vec{i}\in \Lambda}\left(X^{n_{\vec{i}}} Z^{m_{\vec{i}}}\right).
\end{equation}
Here $X_\vec{i}$, $Y_\vec{i}$, $Z_\vec{i}$ denote the Pauli matrices on site $\vec{i}$, $\phi,\ n_\vec{i},\ m_\vec{i}=0.1$. 
The group $\mathcal{G}_\Psi$ is abelian, and generated by $N$ independent Pauli strings, denoted $\hat{g}_\mu$.  The state is uniquely identified and corresponds to
\begin{equation}
    \rho = |\Psi\rangle\langle \Psi| = \prod_{\mu=1}^N \left(\frac{1+\hat{g}_\mu}{2}\right) = \frac{1}{2^N}\sum_{g\in \mathcal{G}_\Psi}g.\label{eq:defstate}
\end{equation}
Additionally, we shall consider \emph{mixed} stabilizer states defined as follows~\cite{aaronson2004improved}. Consider the stabilizing group $|\mathcal{G}_\rho|=2^r$.  We denote as stabilizer state also 
\begin{equation}
    \rho = \frac{1}{2^N}\prod_{\mu=1}^r \left({1+\hat{g}_\mu}\right)\label{eq:defstate2}.
\end{equation}
As we detail, this choice is motivated by both Eqs.~\eqref{eq:defstate}-\eqref{eq:defstate2} mapping to new stabilizer states through the stabilizer circuits below. 

Stabilizer circuits are quantum circuits with two building blocks: (i) Clifford unitary gates and (ii) projective measurements on a Pauli string of the form Eq.~\eqref{eq:defpauli}. Clifford unitary gates are a subgroup of the full unitary group that maps a Pauli string to a \emph{unique} Pauli string. Consequently, a stabilizer state with stabilizer group $\mathcal{G}$  is mapped through Clifford gates $U$ to a new stabilizer state with stabilizer group $U\mathcal{G}U^\dagger $. (For a pure state, this follows from $|\Psi'\rangle \equiv U |\Psi\rangle = U g U^\dagger U |\Psi\rangle \equiv g' |\Psi'\rangle$; identical considerations apply for mixed stabilizer states Eq.~\eqref{eq:defstate2}.)

Similarly, projective measurements on a Pauli string $g^\star$ map a stabilizer state to a new stabilizer state. If the stabilizer group uniquely determine the state (equivalently, the state is pure), then either $[g^\star,\hat{g}_\mu]=0$ for all generators, or there exists a $\hat{g}_\alpha$ such that $[ g^\star, \hat{g}_\alpha]\neq0$~\footnote{This is the most general scenario for qubits. Indeed, if there exists a set
$\{g_{{\alpha}}\}_{\alpha=\alpha_1\dots\alpha_b}$ such that $[g_{\alpha_k},g^\star]\neq0$ for all $k=1,\dots,b$, we can consider the choice of generators $\{g_{\alpha_1}\}\cup \{ g_{{\alpha_1}}\cdot g_{{\alpha_k}}\}_{k>1}$. From the Pauli algebra, it follows the second set $\{g_{{\alpha_1}}\cdot g_{{\alpha_k}}\}$ commute with $g^\star$.  }. 
In the first case either $g^\star$ or $-g^\star$ is in $\mathcal{G}_\Psi$, hence the state is left unchanged by the projection, and the measurement result is deterministic~\footnote{The latter can be obtained by solving $g^\star \hat{g}_\mu g^\star = \hat{g}_\star$ for an indeterminate global phase $\phi$, which in turn is a linear equation in the field $\mathbb{F}_2$.}.
In the second case, the measured string does not belong to the stabilizer group $\mathcal{G}_\Psi$.
The measurement result is random $\pm1$, and the state collapses to the resulting Pauli string $\pm g^\star$. The post-measurement stabilizer group $\mathcal{G}_\Psi'$ is spanned by the same generators of $\mathcal{G}_\Psi$, but with $\hat{g}_\alpha$ is replaced with $\pm g^\star$ depending on the measurement result. 

The above statements are the Gottesman-Knill theorem. It follows that numerical simulations are performable in polynomial time. Indeed, if the stabilizer state at the initial time is representable by a $N\times(2N+1)$ tableau $G$, whose rows are 
\begin{equation}
\label{eq:tableau_def}
    G_{\mu} = (\phi^\mu,\{n^\mu_\vec{i}\},\{m^\mu_\vec{i}\})
\end{equation}
then Clifford gates and projective measurements map a tableau to a new tableau $G'$ through $\mathbb{F}_2$ operations.

Analogous ideas apply if the stabilizer state $\rho$ is not pure ($r<N$). The only differences are that the tableau is  $r\times (2N+1)$, and the measurement operation can now increase the rank of the stabilizer when $\pm g^\star$ commutes with all the elements in $\mathcal{G}_\rho$ but is not in $\mathcal{G}_\rho$. 
In this case, the measurement result is random, and $\pm g^\star$ is a new row in the tableau, which now becomes $(r+1)\times (2N+1)$.

\subsection{Circuit architecture}
In the remaining of this paper, we discuss the measurement-induced phase transition in $(d+1)$-dimensional stabilizer circuits. Here we motivate and specify the choice of circuital architecture in the main text. Nevertheless, we anticipate our results are universal and do not depend on the microphysics of the system (see the further benchmarks in Appendix~\ref{app:tests}).

\paragraph{One-dimensional stabilizer circuits}
For (1+1)-dimensional circuits we consider the regular brick-wall structure in Ref.~\cite{li2018quantum,li2019measurementdriven}. A single evolution step comprises:: (i) a measurement layer stochastically projecting the system qubits and (ii) a unitary layer of Clifford gates. 
The former is defined by
\begin{equation}
\begin{split}
    |\Psi\rangle \mapsto |\Psi'\rangle = \frac{\prod_{\mathbf{i}\in \Lambda} M_{\mathbf{i}}^{h_i}|\Psi\rangle}{\|\prod_{\mathbf{i}\in \Lambda} M_{\mathbf{i}}^{h_i}|\Psi\rangle\|},\label{eq:measlayer} \\
    M^0_\mathbf{i} = \openone_\mathbf{i},\quad M^\pm_\mathbf{i} = \frac{\openone_\mathbf{i}\pm Z_\mathbf{i}}{2},
\end{split}
\end{equation}
where ${h_i=0,+,-}$ is a stochastic variable with probabilities ${\mathcal{P}_0=(1-p)},\mathcal{P}_\pm=p|\langle \Psi|M^\pm_\mathbf{i}|\Psi\rangle|^2$.
Conversely, the unitary layers are given by
\begin{equation}
    U(t) = \prod_{x=1}^{L/2} U_{2x-\mathrm{mod}(t,2),2x+1-\mathrm{mod}(t,2),t},
    \label{eq:uni1d}
\end{equation}
where $\mathrm{mod}(a,b)$ is $a$ modulo $b$. Eq.~\eqref{eq:uni1d} is build from the two-body random Clifford gates $U_{\vec{x},\vec{y},t}$, acting on the qubits $\vec{x}, \vec{y}$ at time $t$.

\paragraph{Two and three-dimensional circuits}
In $d\ge 2$, we consider a $(d+1)$-dimensional randomized stabilizer circuit. This setup is defined by alternating the measurement layers in Eq.~\eqref{eq:measlayer} to a randomized unitary layer, with nearest-neighboring two body gates randomly sampled from the Clifford group throughout the lattice
\begin{align}
    U(t) = \prod_{\langle \vec{x},\vec{y}\rangle \in I_t} U_{\vec{x},\vec{y},t}
    \label{eq:randomized}
\end{align}
with $I_t$ the sample of random neighboring sites on the $d$-dimensional lattice $\Lambda$. Each set $I_t$
consists of $|\Lambda|/4$ pairs of points $\langle \vec{x},\vec{y}\rangle$ chosen in the following manner: first, we choose $|\Lambda|/4$ non-overlapping points $\vec{x}$ uniformly from the whole lattice $\Lambda$, then for each of the points we select $\vec{y}$ as a neighboring site in one out of $d$ randomly chosen directions.

Compared to the structured architectures (e.g., for (2+1)D stabilizer circuits in Ref.~\cite{turkeshi2020measurementinduced,lunt2021measurementinduced}), the randomized circuit simplifies analysis of numerical results as it is invariant (upon the averaging) with respect to translations of time $t \to t+1$, whereas the structured architectures have smaller symmetry $t \to t+T$ with $T>2$ (for instance $T=4$ for the circuit used in \cite{turkeshi2020measurementinduced}). 
Nevertheless, in Appendix~\ref{app:tests} for completeness and comparison with the literature for (2+1)D, we detail the numerical results for (i) the plaquette structure in Ref.~\cite{turkeshi2020measurementinduced} and (ii) the brick-wall structure in Ref.~\cite{lunt2021measurementinduced}. All these frameworks give consistent and equivalent results, as expected by universality.
The (3+1)D setup is a crucial novelty of our work, extending the investigation of measurement-induced phase transition to $d=3$ spatial dimensions. To test the universality of our results, we report in Appendix~\ref{app:tests} numerical simulations using a cubic regular pattern.

\begin{figure*}[t!]
    \centering
    \includegraphics[width=\textwidth]{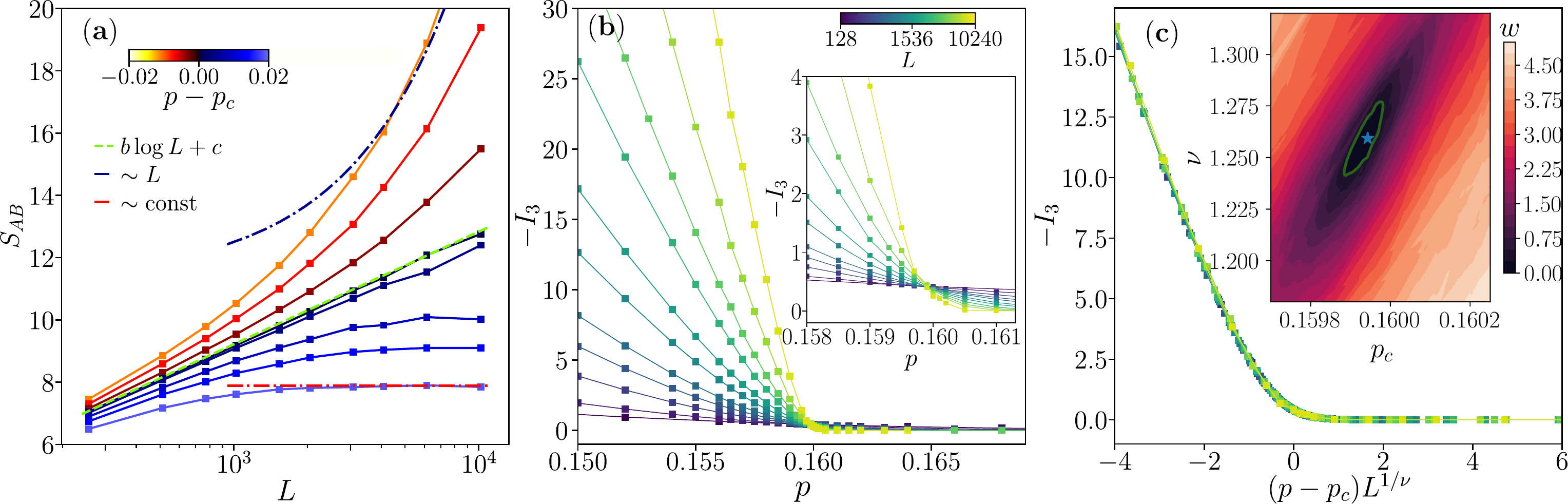}
    \caption{Entanglement transition in (1+1)D circuits. (a) Half-system entanglement entropy $S_{AB}$ as function of system size $L$ for various measurement rates $p$ (color-coded, changing from $p=0.1585$ to $p=0.162$). Below the transition ($p<p_c$) the entropy grows linearly with system size (volume-law), while above ($p>p_c$), it saturates to a constant (area-law). At the transition point, $p_c=0.15995(10)$, the scaling is logarithmic, $S_{AB} =c_{\mathrm{eff}}  \ln L +c$, with $c_{\mathrm{eff}} =1.57(1)$ and $c=-1.62(1)$. (b) TQMI, $I_3$, for various $L$ (color-coded, from $L=128$ to $L=10240$) and $p$. Inset: a close-up around the crossing point where the transition is located. (c) The finite-size scaling collapse for $p_c=0.15995(10)$ and  $\nu=1.260(15)$. These values are obtained by minimizing the cost function $\mathcal{W}(p_c,\nu)$ as described in Appendix~\ref{app:ffs}. Inset: the value of $w=\ln \left( \mathcal{W}(p_c,\nu)/\mathcal{W}(p^\star_c,\nu^\star)\right)$ across the parameter space, with darker colors locating regions closer to the minimum $(p_c^\star,\nu^\star)$ (denoted by the blue star). The green line bounds the region $\mathcal{W}(p_c,\nu)\le 1.3\mathcal{W}(p_c^\star,\nu^\star)$.}
    \label{fig:1dimage}
\end{figure*}

\begin{figure*}[t]
    \centering
    \includegraphics[width=\textwidth]{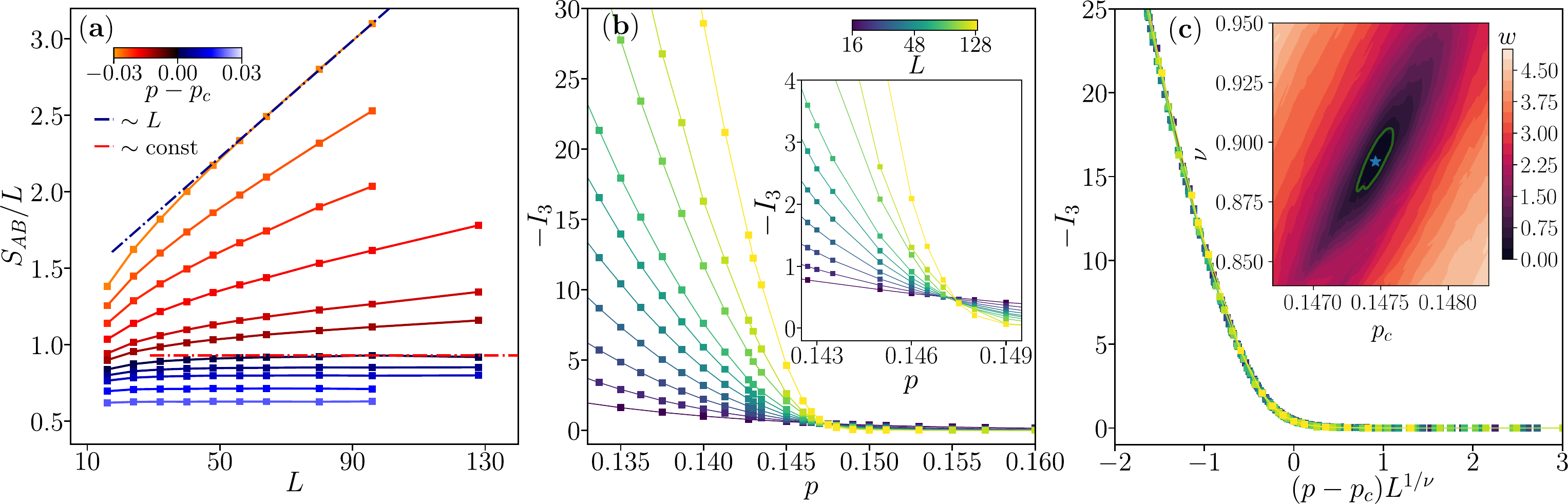}
    \caption{Entanglement transition in (2+1)D circuits. (a) Scaling of the half-system entanglement entropy $S_{AB}$  for various system sizes $L$ and measurement rates $p$ (color-coded, from $p=0.12$ to $p=0.175$).  Below the critical point, at $p<p_c=0.1475(2)$, we observe a volume-law $S_{AB} \sim L^2$ , while for $p>p_c$ an area law, $S_{AB} \sim L$, is obeyed. (b) TQMI, $I_3$, for various $p$ and $L$ (color-coded, from $L=16$ to $L=128$). Inset: close-up around the crossing point, where the transition is located. 
    (c) Data collapse for the critical parameters $p_c=0.1475(2)$, $\nu=0.889(17)$. Inset: landscape of the cost function $w$ discussed in Appendix~\ref{app:ffs}.} 
    \label{fig:2dimage}
\end{figure*}

\begin{figure*}[t!]
    \centering
    \includegraphics[width=\linewidth]{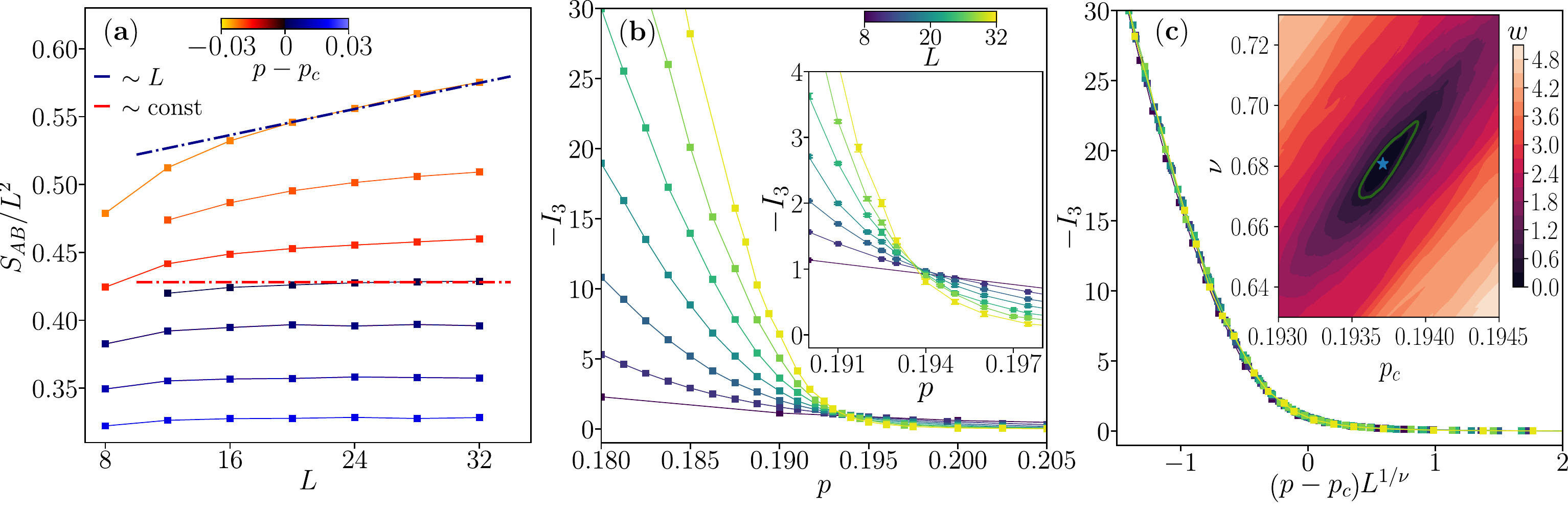}
    \caption{Entanglement transition in (3+1)D circuits. (a) Scaling of the half-system entanglement entropy $S_{AB}$ for various system sizes $L$ and measurement rates $p$ (color-coded, from $p=0.18$ to $p=0.22$).  Below the critical point, at $p<p_c=0.1937(2)$, we observe a volume-law $S_{AB} \sim L^3$, while for $p>p_c$ an area law, $S_{AB} \sim L^2$, is obeyed. (b) TQMI, $I_3$, for various $p$ and $L$ (color-coded, from $L=8$ to $L=32$). Inset: close-up around the phase transition. (c) Data collapse for the critical parameters $p_c=0.1937(2)$, $\nu=0.681(20)$. Inset: landscape of the cost function $w$ (cf. Appendix~\ref{app:ffs}) used to locate the critical parameters. }
    \label{fig:3dimage}
\end{figure*}

\section{Entanglement transition}
\label{sec:results}
We fix $\Lambda$ to be a hypercube of linear dimension $L$ with periodic boundary conditions in every direction (see Fig.~\ref{fig:2dimage}). We start from the unpolarized pure state $|\Psi_0\rangle = |0\rangle^{|\Lambda|}$ and evolve the circuit up to time $t=10L$, after which the system reaches the stationary state (in which the average value of the entanglement entropy and of the other observables are time-independent).
We compute the half-system entanglement entropy and the tripartite mutual information with $|A|=|B|=|C|=|D|$ being the subsequent connected regions, as shown in Fig.~\ref{fig:partition}. 

For stabilizer states, the entanglement entropy can be easily computed and is given by~\cite{hamma2004,Hamma_2005}
\begin{equation}
    S_A(\rho) = |A| - \log_2 |\mathcal{G}_A|,
\end{equation}
where $\mathcal{G}_A$ is the subgroup of all elements of $\mathcal{G}_\Psi$ that act trivially on the complementary $A_c$. (That is $g \in \mathcal{G}_\Psi$ is such that $g=\tilde{g}_A\otimes \mathbb{I}_{A_c}$ for some non-trivial Pauli string $\tilde{g}_A$ acting on the qubits in $A$). 
Performing some algebraic manipulation one gets the computationally simpler formula~\cite{nahum2017quantum}
\begin{equation}
\begin{split}
    S_A(\rho) & = \mathit{rk}_{\mathbb{F}_2}(G_A) -  |A|\\ 
    G_A&\equiv [\{n^\mu_\vec{i}\},\{m^\mu_\vec{i}\}]_{\mu=1,\dots,|\Lambda|,\vec{i}\in A}
\end{split}
\end{equation}
where $\mathit{rk}_{\mathbb{F}_2}(\cdot)$ denotes the matrix rank on the Galois field $\mathbb{F}_2$ and $G_A$ is the submatrix of $G$ (cf. Eq.~\eqref{eq:tableau_def}) without phases $\phi^{\mu}$ and with lattice indices restricted to the subsystem $A$.

Here, we shall consider $S_{AB}=\mathbb{E}_\mathbf{m}[S_{AB}(\rho_\mathbf{m})]$ in the stationary state, where $AB\equiv A\cup B$ is the half of the system. Similarily, we  consider the average TQMI, $I_3$ \eqref{eq:i3}.
The finite size scaling analysis in this section and in the following ones are performed by cost function optimization, as detailed in Appendix~\ref{app:ffs}.

\paragraph{One-dimensional stabilizer circuits}
We investigate the tripartite mutual information with a quadripartition of the system into intervals of equal length $L/4$, as shown in  Fig.~\ref{fig:partition}(a). Calculating the entanglement entropy $S_{AB}$ as a function of the system size $L$, we observe a qualitative change in its system size dependence from the volume-law $S_{AB} \sim L$ to area-law  $S_{AB} \sim \mathrm{const}$, as the measurement rate $p$ is increased, see Fig.~\ref{fig:1dimage}(a). This behavior is the essence of the measurement-induced entanglement transition. Following the reasoning of Sec.~\ref{sec:overview}~\textit{b}, we turn our attention to the average TQMI, and we indeed observe a crossing point at which $I_3 =\mathcal O(1)$ is independent of $L$ around a measurement rate $p_x=0.1600(1)$, see Fig.~\ref{fig:1dimage}(b). The finite-size scaling of data with a power-law divergent correlation length \eqref{eq:I3ansatz} shows that MIPT occurs at that measurement rate. Indeed, using system sizes
up to $L=N=|\Lambda|=10240$, we estimate $p_c=0.15995(10)$ and $\nu=0.1260(15)$, see Fig.~\ref{fig:1dimage}(c) for the collapse of the data and the landscape of the employed cost function $w$. In passing, we note that compatible values of the critical parameters can be obtained by finite system size analysis of entanglement entropy with an ansatz:
\begin{equation}
    S_{AB}(p,L)= S_{AB}(p_c,L) + F'[(p-p_c)L^{1/\nu}],
    \label{eq:Sansatz}
\end{equation}
which, however, leads to larger uncertainties regarding the critical parameters (data not shown).

At the critical point, $p=p_c$, the scaling of the entanglement entropy is logarithmic, $S_{AB} = c_{\mathrm{eff}} \log L + b$, with coefficient $c_\mathrm{eff} = 1.57(1)$, refining previous estimates~\cite{zabalo2020critical}. 
In particular, our analysis confirms the inequivalence of the measurement-induced criticality and the 2D percolation field theory with a larger confidence interval with respect to earlier works. 

\paragraph{Two-dimensional circuits}
We investigate the half-system entropy $S_{AB}$ and tripartite quantum mutual information $I_3$ in a quadripartition of connected cylinders with equal volume $L^2/4$, see Fig.~\ref{fig:partition} (b). Employing the randomized circuital architecture (cf. Eq.~\eqref{eq:randomized}) we reach system sizes up to $L=128$ ($N=16384$ qubits).

Investigating system size dependence of $S_{AB}$ for various measurement rates $p$, we observe a change between a volume-law scaling of entanglement entropy $S_{AB} \sim L^2$ and an area-law behavior $S_{AB} \sim L$, see Fig.~\ref{fig:2dimage}(a). This is a signature of MIPT in the considered system. In order to locate the transition, we calculate TQMI, $I_3$, and observe a crossing point at $p_x= 0.1475(5)$, which separates a regime with behavior $I_3 \sim L^2 $ from a regime in which TQMI decays exponentially with system size, see Fig.~\ref{fig:2dimage}(b). The two regimes can be identified with QEC and QZ phases as the finite-size scaling of $I_3$ with the ansatz \eqref{eq:I3ansatz} shows, cf. Fig~\ref{fig:2dimage}(c). We obtain the critical measurement rate and the exponent
\begin{equation}
    p_c=0.1475(1),\qquad \nu =0.887(20).
    \label{eq:res2drand}
\end{equation}
The finite size analysis of data for $I_3$ was performed for system sizes $40\leq L \leq 128$. A similar scaling analysis of entanglement entropy $S_{AB}$, according to the ansatz in Eq.\eqref{eq:Sansatz} is associated with larger finite size effects. Indeed, we observe a monotonous flow of the exponent $\nu$ in such an analysis: collapsing data for $S_{AB}$ for $L \leq 64 $ leads to the exponent $\nu\approx0.74$, whereas such an analysis for system sizes $L\geq64$ yields $\nu\approx0.82$. The TQMI is not associated with such strong finite system-size  drifts. This explains the discrepancy between the estimate of $\nu$ from Ref.~\cite{turkeshi2020measurementinduced} based on the scaling of half-system entanglement entropy and the value of $\nu$ obtained in the present work which is compatible with the results of  Ref.~\cite{lunt2021measurementinduced}.

To illustrate the universal behavior at the entanglement phase transition, we report here the main results of Appendix~\ref{app:tests}. 
The critical measurement rate for the randomized circuit architecture is considerably smaller than for the plaquette ($p^\mathrm{plaq}_c=0.5517(1)$)  and brick-wall frameworks ($p^\mathrm{brick}_c=0.3116(1)$), while the critical exponents $\nu$ describing the power-law diverging lengthscale at the transition are all compatible among themselves ($\nu^\mathrm{plaq}=0.839(22)$, $\nu^\mathrm{brick}=0.863(16)$ ). 

We stress that Eq.~\eqref{eq:res2drand} is compatible with the correlation length critical exponent of the  3D percolation field theory~\cite{phifiveloops}. Moreover, as we shall see, this compatibility between (2+1)D measurement-induced criticality and 3D percolation field theory will extend to all the critical exponents (see Table~\ref{tab:critical_exponents}). 
Lastly, at the critical point, $p=p_c$, the entanglement entropy $S_{AB}$ scales linearly with the system size, in alignment with a 3D CFT~\cite{Calabrese_2004} and in contrast with the behavior reported in Ref.~\cite{turkeshi2020measurementinduced}. As noted in Ref.~\cite{lunt2021measurementinduced}, this discrepancy is due to the large finite size effects present in the entanglement entropy that require larger system sizes for proper identification of the critical point.

\paragraph{Three-dimensional circuits}
We consider the randomized circuits (cf. Eq.~\eqref{eq:randomized}) in (3+1) spatiotemporal dimensions and compute the entanglement entropy scaling and the TQMI for a quadripartition of the system into equal hyper-cylinders, each with volume $L^3/4$, see Fig.~\ref{fig:partition}(c). 

Calculation of the half-system entanglement entropy for increasing measurement rate $p$ reveals a crossover between a volume-law regime with $S_{AB}\sim L^3$ and an area-law regime $S_{AB}\sim L^2$, as shown in Fig.~\ref{fig:3dimage}(a). This is the basic feature of the measurement-induced entanglement transition in random circuits in $d=3$ spatial dimensions. To investigate the MIPT in more detail, we calculate TQMI and find a clear crossing point of curves $I_3(p)$ which separates regimes with extensive and exponentially decaying system size dependence of $I_3$, as shown in Fig.~\ref{fig:3dimage}(b).
Performing the finite-size scaling with Eq.~\eqref{eq:I3ansatz}, we find a very good collapse of the data, see Fig.~\ref{fig:3dimage}, indicating a presence of entanglement transition in the (3+1)D circuits. Optimization of the collapse yields the following critical parameters
\begin{equation}
    p_c = 0.1937(1),\qquad \nu=0.686(20),\label{eq:3dest}
\end{equation}
with the correlation length critical exponent compatible with that of 4D percolation~\cite{phifiveloops}, see Table~\ref{tab:critical_exponents}.
As for the two-dimensional case, we present evidence of universal behavior by comparing Eq.~\eqref{eq:3dest} with the results in Appendix~\ref{app:tests} for the cubic structured pattern ($p_c^\mathrm{cubic}=0.7814(4)$, $\nu^\mathrm{cubic}=0.662(22)$).
In this case, the critical measurement rate is lower than in the randomized setup, while the correlation length critical exponents $\nu$ are mutually compatible within one error bar. 
In line with a conformal field theory in 4D, the entanglement entropy scales as $L^2$ at the critical point.

\begin{figure*}[t!]
    \centering
    \includegraphics[width=0.98\linewidth]{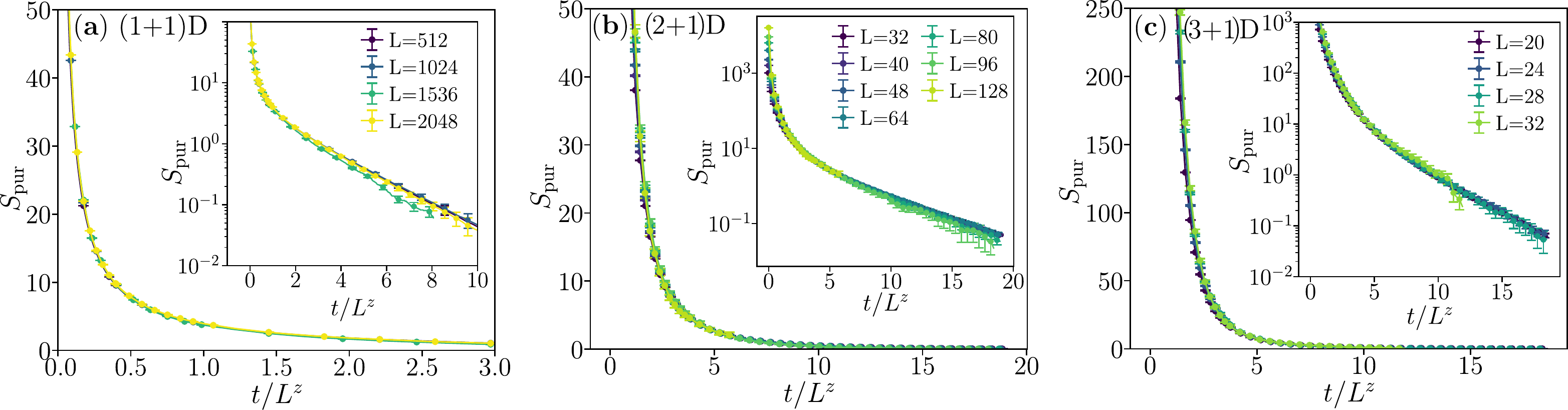}
    \caption{Purification transition: entropy $S_{\mathrm{pur}}$ of the initially maximally mixed state as a function of $t/L^z$, where $L$ is the system size, and $z$ is the estimated dynamical exponent.  Data for (\textbf{a}) (1+1)D circuits; (\textbf{b}) (2+1)D circuits, and (\textbf{c}) (3+1)D circuits. By optimizing the collapse of the curves (taking into account times $t$ for which $S_{\mathrm{pur}}\leq 50$ for $d=1,2$ and $S_{\mathrm{pur}}\leq 250$ for $d=3$), we obtain $z=1.00(1)$ for (1+1)D; $z=1.01(2)$ for (2+1)D; $z=1.02(4)$ for (3+1)D. In the insets, we plot the same data in a linear-logarithmic scale to highlight the decay of $S_{\mathrm{pur}}$ at late times. }
    \label{fig:z}
\end{figure*}

\begin{figure}[t!]
    \centering
    \includegraphics[width=\columnwidth]{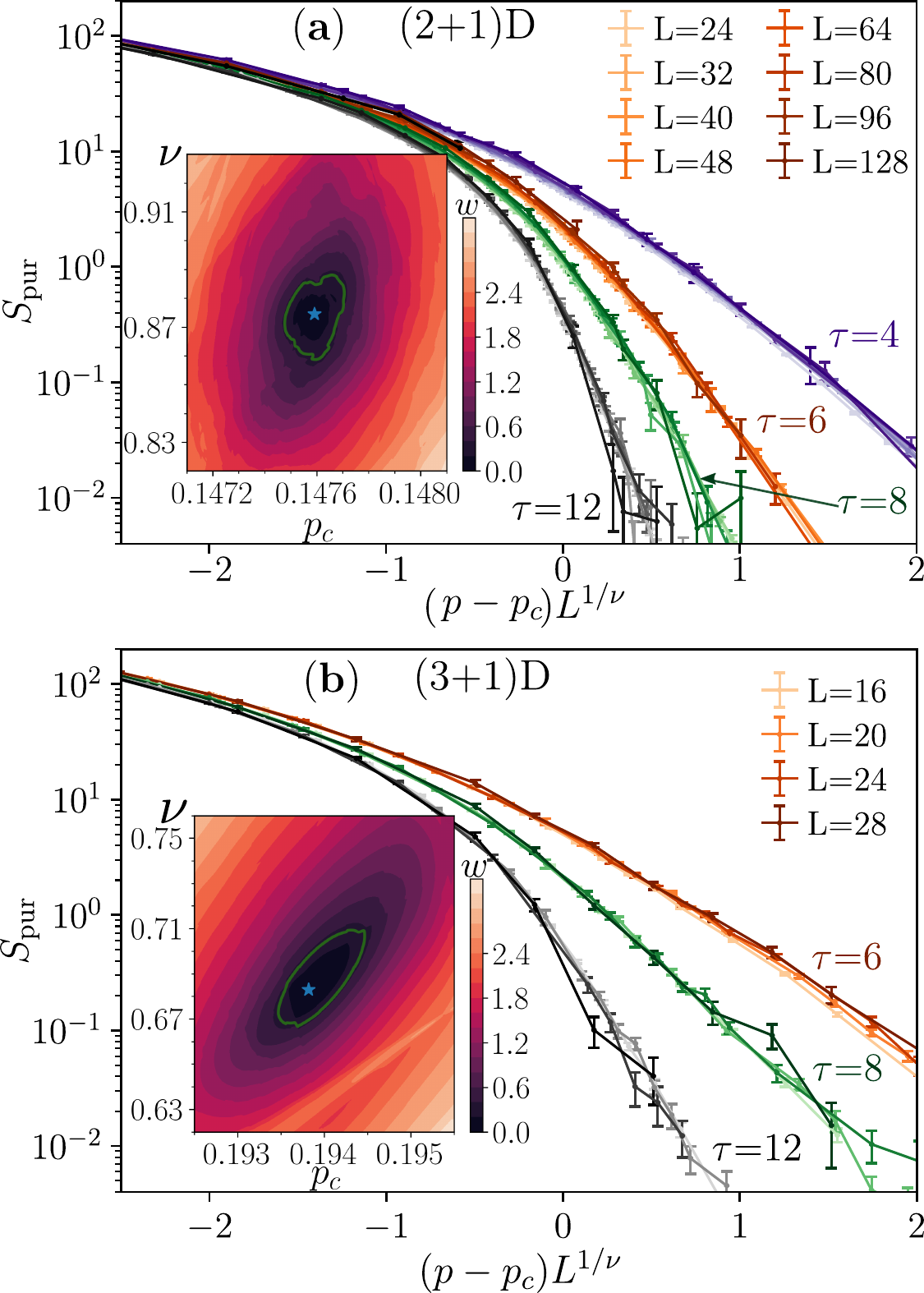}
    \caption{Purification transition: collapse of the entropy $S_{\mathrm{pur}}$ at various rescaled times $\tau = t/L$ for (2+1)D (\textbf{a}) and  (3+1)D  (\textbf{b}) circuits. The insets show the landscape of the cost function $w$ (cf. Appendix~\ref{app:ffs}) used to locate the critical parameters: $p_c = 0.1476(1)$, $\nu = 0.875(15) $ for the (2+1)D case and $p_c = 0.1937(5)$, $\nu = 0.68(3)$ for (3+1)D circuits.  }
    \label{fig:puriCOL}
\end{figure}

\section{Purification transition}
\label{sec:purtrans}
This section studies the purification transition in the stabilizer circuits. We consider a $d$-dimensional hypercubic lattice $\Lambda$ with periodic boundary conditions in every direction, initialize the system in a fully mixed state $\rho_0\propto \openone$, and let the system evolve under the circuit architectures described in Sec.~\ref{sec:methods}. The observable of interest is the entropy of the state of the system. The entropy of a mixed stabilizer state $\rho$, determined by a set of $r<N=|\Lambda|$ independent generating Pauli strings $\hat g_{\mu}$ (cf. Eq.~\eqref{eq:defstate2}), 
is given by the key result~\cite{hamma2004,Hamma_2005}
\begin{equation}
    S(\rho) = |\Lambda| - r = N-r. \label{eq:enttherm}
\end{equation}
Below, we study the time evolution of the entropy averaged over quantum trajectories $S_\mathrm{pur}=\mathbb{E}_\mathbf{m}[S(\rho_\mathbf{m})]$.

\paragraph{Dynamical critical exponent $z$}

First, we fix the measurement rate as $p=p_c$ (with the critical measurement rate determined in Sec.~\ref{sec:results}) and consider the process of purification of the maximally mixed initial state by investigating the dependence of $S_\mathrm{pur}$ on time (circuit depth) $t$. Accordingly with our expectations, we find a monotonous decrease of $S_\mathrm{pur}$ as a function of time $t$ that occurs in such a way that $S_\mathrm{pur}$ plotted as a function of $t/L^z$ collapse onto universal, system size independent curves. We determine the values of the dynamical critical exponent $z$ by optimizing the collapse, as shown in Fig.~\ref{fig:z}, which yields 
\begin{align}
    d=1 & \qquad z=1.00(1),\nonumber\\
    d=2 & \qquad z=1.01(2),\label{eq:resz}\\
    d=3 & \qquad z=1.02(4).\nonumber
\end{align}
Due to larger system sizes considered, Eq.~\eqref{eq:resz} overcomes previous numerical computations in Ref.~\cite{zabalo2020critical} for (1+1)D stabilizer circuits and Ref.~\cite{lunt2021measurementinduced} for (2+1)D stabilizer circuits, as well as shows that the dynamical critical exponent $z$ is compatible with unity for $d=3$. In particular, this value of the exponent $z$ is consistent with the scaling invariance at the critical point for $d=1,2,3$, in alignment with its conformal field theory description. 

\paragraph{Purification phase transition for $d=2,3$}

Here, we use the estimated dynamical exponent $z=1$ to extract the correlation length critical exponents for the purification dynamics in (2+1)D and (3+1)D stabilizer circuits. To this end, following the ansatz \eqref{eq:pur}, we study the system size scaling of the average entropy $S_\mathrm{pur}$ close to the critical point for a fixed value of $\tau\equiv t/L$. By optimizing the collapse, we find
\begin{equation}
\begin{split}
    d=2 & \qquad \nu^\mathrm{pur}=0.875(15),\\
    d=3 & \qquad \nu^\mathrm{pur}=0.68(3),
    \end{split}
    \label{eq:nupur}
\end{equation}
with the data collapses given in Fig.~\ref{fig:puriCOL}(a) and (b) respectively for (2+1)D and (3+1)D stabilizer circuits. 

Our study of the purification phase transition employed the critical measurement rate $p_c$ from Sec.~\ref{sec:results} to obtain the value of the exponent $z$. We could have disposed of this dependence by considering a generalization of the ansatz in \eqref{eq:pur}, where $t/L$ is substituted by $t/L^z$. Collapsing $S_\mathrm{pur}$ according to such a generalized ansatz would result in an independent determination of the three critical parameters: $p_c$, $\nu$, and $z$. However, such a scaling analysis would be less constrained, leading to larger uncertainties in the estimated critical parameters. This is not desirable, especially for the $d=3$ case, in which the range of available system sizes is limited. For that reason, we have
opted for the ansatz \eqref{eq:pur} which already assumes $z=1$.
Nevertheless, our numerical findings for the entanglement phase transition (cf. Sec.~\ref{sec:results}) are in full compatibility with the critical measurement rate and correlation length critical exponent found in the investigation of the purification measurement-induced transitions. This shows that the two transitions are two aspects of the same physical phenomenon, the measurement induced criticality, not only for $d=1$ but also for $d=2,3$.

\begin{figure*}[t!]
    \centering
    \includegraphics[width=0.867\columnwidth]{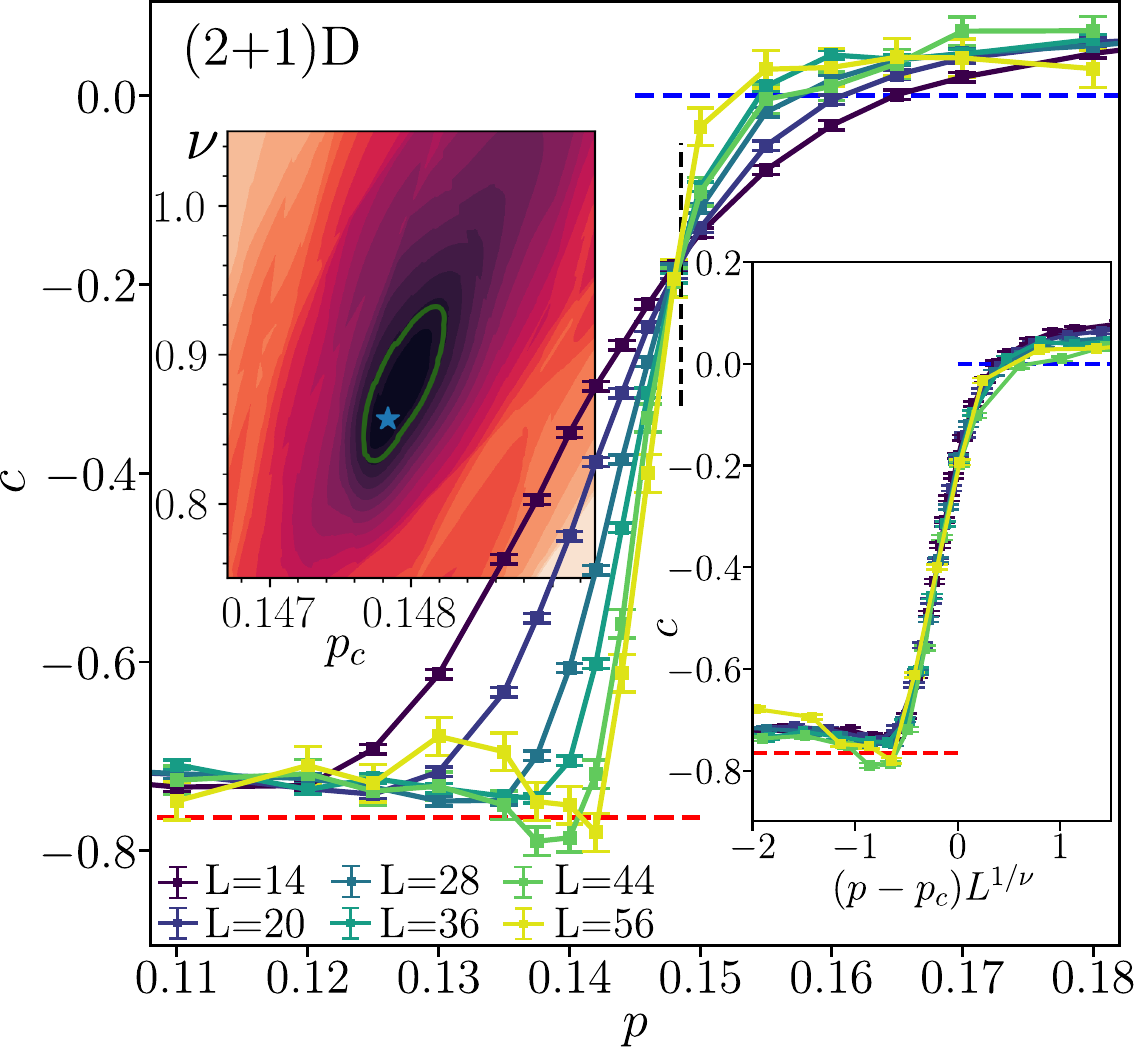}\includegraphics[width=0.85\columnwidth]{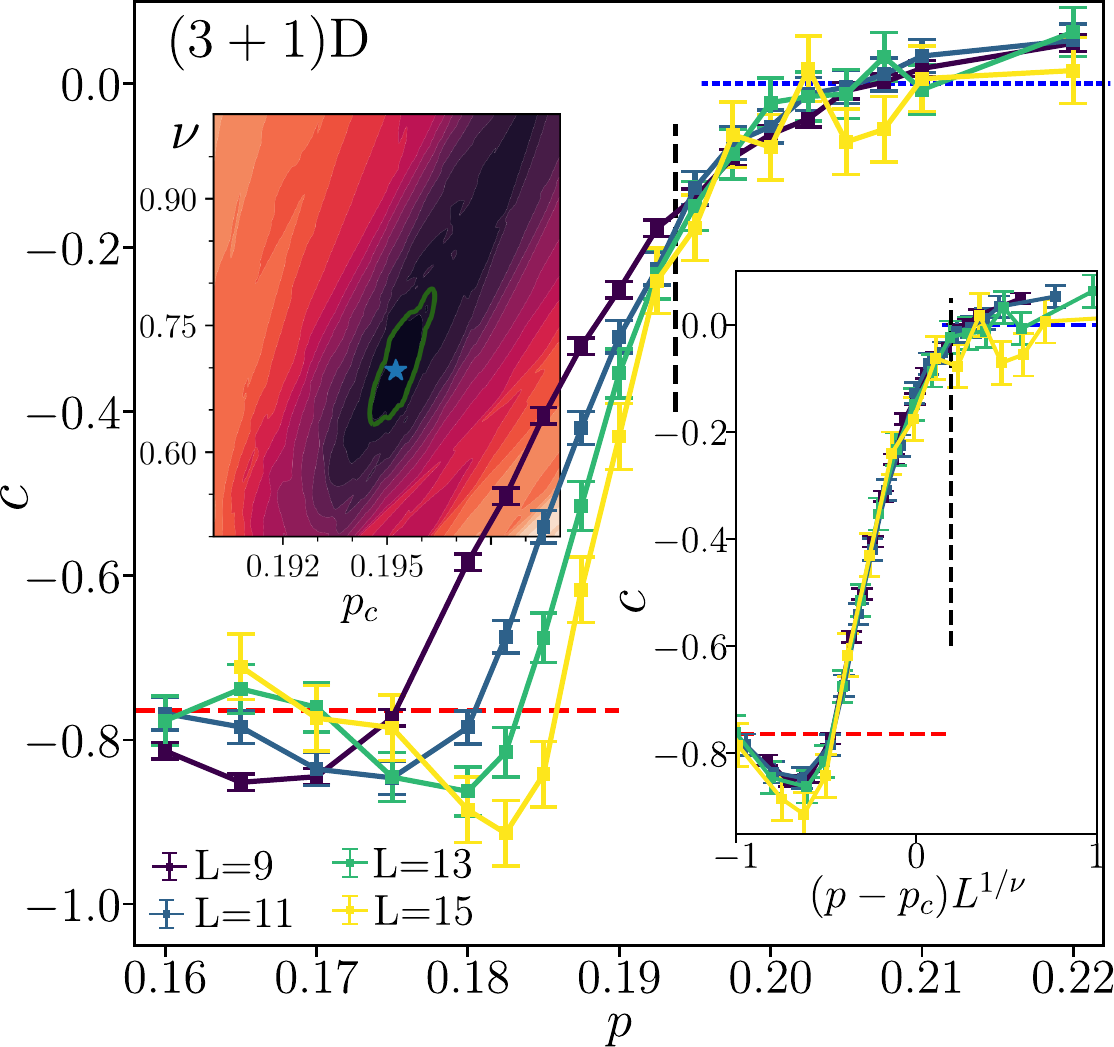}
    \caption{Sub-leading term $c$ in system size scaling of participation entropy $S^{\mathrm{part}} = D |\Lambda| + c$ as function of measurement rate $p$ at MIPT in (2+1)D (left panel) and (3+1)D (right panel) circuits for various system sizes $L$. The red and blue dashed lines show the analytical predictions for $p\to0$ limit ($c\simeq -0.7645$, see \eqref{eq:cexact}) and for $p\to1$ limit ($c=0$).
    The color maps in insets show the landscapes of the cost function $w$ (cf. Appendix~\ref{app:ffs}) used to locate the critical parameters: $p_c = 0.1478(6)$, $\nu = 0.86(6) $ for the (2+1)D case and $p_c = 0.195(2)$, $\nu = 0.69(7)$ for (3+1)D circuits.  }
    \label{fig:cq2D}
\end{figure*}

\section{Structure of wave-functions}
\label{sec:struct}
This section discusses the change in properties of wave functions along quantum trajectories across the MIPT. We investigate the participation entropy of a pure state $\rho=|\Psi\rangle\langle \Psi|$ obtained by application of the stabilizer circuits introduced in Sec.~\ref{sec:methods}, initialized in state $\rho_0=|0\rangle^{\otimes N}$. We consider a steady-state regime, $t \geq 10L$, in which the considered observables attain time-independent average values. 

\paragraph{Participation entropy and fractal coefficients}
We consider the participation entropy \eqref{sqdef} in the eigenbasis of $Z_i$ operators
which can readily be calculated for a stabilizer state \eqref{eq:defstate} and is given by (see ~\cite{sierant2022universal})
\begin{equation}
    S_\mathrm{part}(\rho) =  \textit{rk}_{\mathbb{F}_2} M_X,
\end{equation}
where $M_X=(m_\mu^\vec{i})$ is the sub-matrix of the $X$-operators in Eq.~\eqref{eq:tableau_def}, and $\textit{rk}_{\mathbb{F}_2}(M)$ is the rank of the matrix $M$ on the field $\mathbb{F}_2$.

We consider the average participation entropy $S_\mathrm{part} = \mathbb{E}_\mathbf{m}[S_\mathrm{part}(\rho_\mathbf{m})]$ in stabilizer circuits and focus on the fractal sub-leading coefficient $c$ in the system size dependence of $S_\mathrm{part}$.
At the level of the statistical field theory for random circuits~\cite{sierant2022universal}, the sub-leading coefficient $c$ is a ferromagnetic-like order parameter, being zero in the QZ phase, at $p>p_c$,  and a non-zero constant
\begin{equation}
    c=1-\frac{1}{\ln 2}\psi_{1/2}\left(1-\frac{i\pi}{\ln 2}\right) \simeq -0.7645,\label{eq:cexact}
\end{equation}
in the QEC phase, for $p<p_c$, where  $\psi_q$ is the $q-$th digamma function~\cite{qpoly}. We note that the non-vanishing value of the sub-leading term $c$ may be interpreted as an excluded volume-effect in single particle systems \cite{sierant2022universality}, and it characterizes also eigenstates of random matrices \cite{Backer19}. Also, we would like to emphasize that while the fractal dimension $D$ is basis-dependent, the sub-leading coefficient $c$ remains invariant under local changes of basis and encodes universal information on the measurement-induced criticality. This was demonstrated in Ref.~\cite{sierant2022universal} for $d=1$ dimensional  quantum circuits. In the following, we show that this hold also for $d=2$ and $d=3$.

Our system is a $d$-dimensional hypercubic lattice $\Lambda$ with periodic boundary conditions, hence Eq.~\eqref{eq:partdef} becomes
\begin{equation}
    S_\mathrm{part} = D L^d+c.
    \label{eq:partdef2}
\end{equation}
In order to extract the fractal dimension $D$ and the sub-leading term $c$ in a system size-resolved manner, one would ideally like to calculate $S_\mathrm{part}$ for two very close system sizes $L$ and $L+\Delta L$ (with $\Delta L \ll L$), and perform a fitting according to \eqref{eq:partdef2}. However, the smaller the value of $\Delta L/L$, the larger the uncertainty of the sub-leading coefficient $c$ obtained in such a fitting procedure. For that reason, in our numerical analysis, we relax the condition $\Delta L \ll L$ in the following way. For (2+1)D stabilizer circuits we increase $\Delta L$ with system size from $\Delta L=4$, for the smallest $L$, up to $\Delta L=16$, for the largest $L$ considered. In turn, for (3+1)D stabilizer circuits we keep $\Delta L=2$, but we are able to investigate a very limited interval of system sizes, up to $L\leq 16$. Finally, we note that when we extract $D$ and $c$ from a fit \eqref{eq:partdef2} to data at system sizes $L$ and $L+\Delta L$, the resulting values are labeled by $L+\frac{\Delta L}{2}$ (and such a value of system size is considered in the finite-system size analysis of the data). In that manner, we extract the fractal dimension $D$ and the sub-leading term $c$ for $d=2,3$. Similarly as for $d=1$~\cite{sierant2022universal}, we observe that $D$ decreases monotonously with the measurement rate $p$ (data not shown), revealing that the wave-functions are fractal in the considered basis for $d=2,3$ and any $p>0$.

\paragraph{The sub-leading term $c$} We present the extracted values of the sub-leading term $c$ in system size scaling of participation entropy in Fig.~\ref{fig:cq2D} for (2+1)D and (3+1)D stabilizer circuits. First, we note that the value of $c$ in the QEC phase, at $p<p_c$, is close to being compatible with the exact result Eq.~\eqref{eq:cexact}. We attribute the small deviations to the condition $\Delta L \ll L$ not being strictly satisfied in our numerical analysis. Similarly, we observe that the sub-leading term $c$, up to small deviations, is vanishing in the QZ regime. Importantly, we observe a crossing point of the curves $c(p)$ at measurement rates that are very close to the critical measurement rates found in Sec.~\ref{sec:results} and~\ref{sec:purtrans} (denoted by black dashed lines in Fig.~\ref{fig:cq2D}). Finally, we perform a finite-size scaling of the data for the sub-leading term $c$, according to the ansatz
\eqref{eq:partcol}, with collapses shown in the insets of Fig.~\ref{fig:cq2D}. This yields the following values of the critical parameters
\begin{align}
    d=2 & \qquad p_c=0.1478(6),\quad \nu=0.86(6),\\
    d=3 & \qquad p_c=0.195(2),\quad\nu=0.69(7).
\end{align}
These results are associated with larger error bars  than the results for the entanglement transition and for the purification transition, mainly
due to the numerical difficulties arising from the fact that $c$ is a \textit{sub-leading} term in system size dependence of participation entropy. Nevertheless, all three approaches (cf. 
Sec.~\ref{sec:results} and~\ref{sec:purtrans})
give results that are compatible within one error bar, providing alternative perspectives on the same measurement-induced criticality.

\begin{figure*}[t!]
    \centering
    \includegraphics[width=\textwidth]{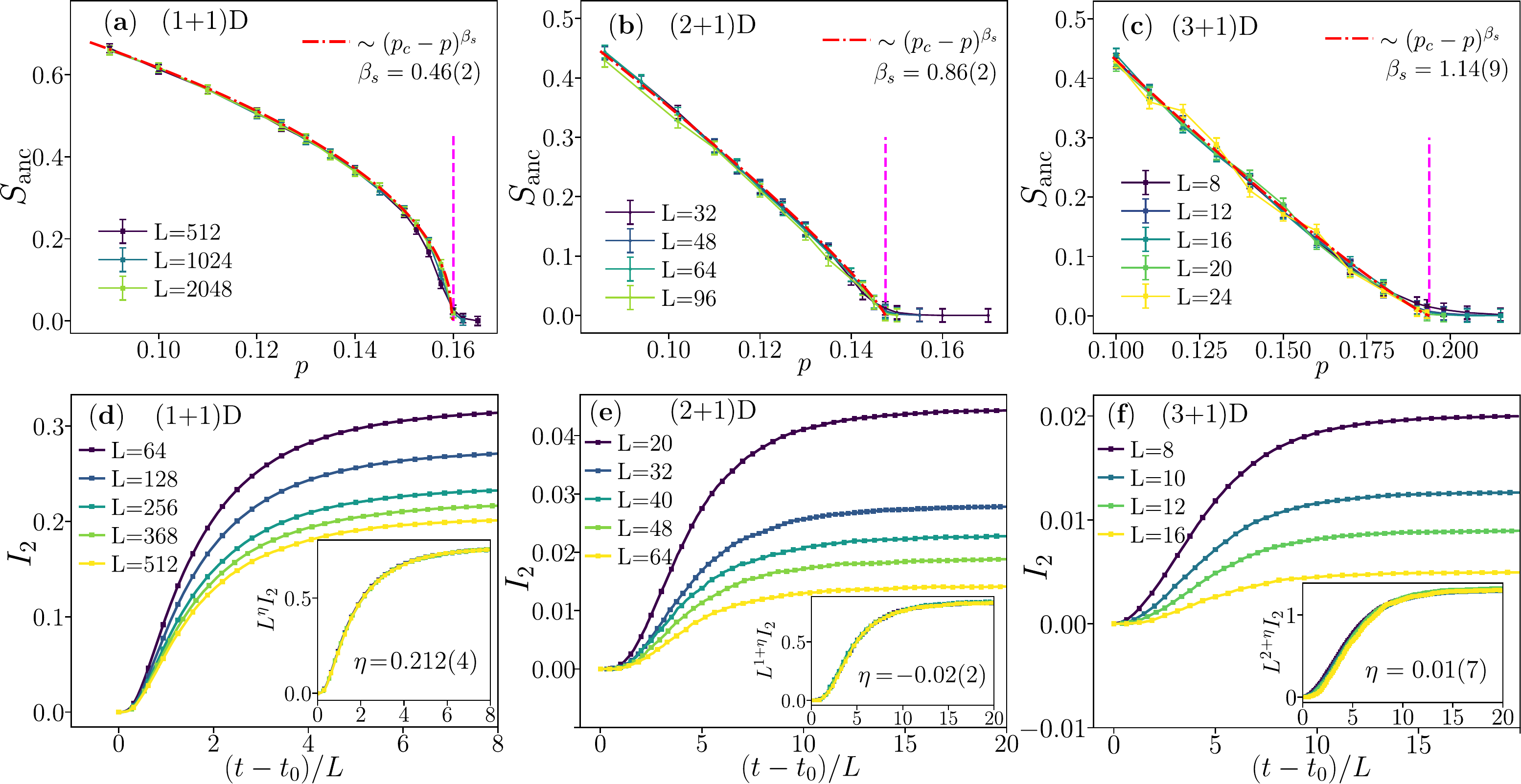}
    \caption{Bulk-surface properties. The upper row: 
    determination of the surface exponent $\beta_s$ which determines the behavior of the ancilla qubit entropy  $S_{\mathrm{anc}}$ as function of the measurement rate $p$: $S_{\mathrm{anc}} \sim (p_c-p)^{\beta_s}$ in (1+1)D (\textbf{a}), (2+1)D (\textbf{b}) and (3+1)D (\textbf{c}) circuits. The magenta dashed lines show the critical measurement rate $p_c$ taken from Sec.~\ref{sec:results}.
    The lower row: quantum mutual information $I_2$ between ancilla qubits separated by distance $L/2$ as a function of the rescaled time $(t-t_0)/L$ (where $t_0=10L$) at measurement rate $p=p_c$ in (1+1)D (\textbf{d}), (2+1)D (\textbf{e}) and (3+1)D (\textbf{f}) circuits. The insets show data collapses 
    that confirm the power-law system size dependence $I_2 \sim  \frac{1}{L^{d-1+\eta}}$ and are used to extract the values of the exponent $\eta$. 
    }
    \label{fig:3deta}
\end{figure*}

\section{Bulk and boundary properties}
\label{sec:blkbnd}

This section investigates the properties of MIPT in $(d+1)$D stabilizer circuits by studying the purification of ancilla qubits~\cite{gullans2020scalable}. We consider a $d$-dimensional hypercubic lattice $\Lambda$ with linear size $L$.
The state of the full system, i.e., of the circuit and of ancilla qubits, is initially given by $|\Phi\rangle \equiv |\Psi\rangle\otimes |a\rangle$, where $|\Psi\rangle$, $|a\rangle$ are the pure states of the circuit and the ancilla. At time $t_0$, the system and the ancilla are entangled through a unitary operation $U$, resulting in a state $|\Phi'\rangle = U|\Phi\rangle$. 
We let the combined system evolve under the quantum circuit, acting non-trivially only on the system qubits
\begin{equation}
     |\Phi_\mathbf{m}\rangle = \frac{K_\mathbf{m}\otimes \openone_a|\Phi'\rangle}{\|K_\mathbf{m}\otimes \openone_a|\Phi'\rangle\|},
     \label{eq:evoANC}
\end{equation}
and compute the ancilla entanglement entropy $S(\rho_{\mathbf{m},a})$ (cf. Eq.~\eqref{eq:entdef}) for $t>t_0$. 
Throughout this section, we are interested in the average residual entanglement entropy
$S_\mathrm{anc}=\mathbb{E}_\mathbf{m}[S(\rho_{\mathbf{m},a})]$.

\paragraph{Bulk and boundary order parameters}
We now focus on studying the bulk and critical boundary exponents,  respectively $\beta$ and $\beta_s$. 
First, we entangle the system and an ancilla at time $t_0$. The full system contains $|\Lambda|+1$ qubits and is in a pure state. Then, we let the full system evolve according to the quantum circuit 
\eqref{eq:evoANC} and compute the entanglement entropy of the ancilla qubit at time $t=t_0+2L$ ($t=t_0+8L$) for $d=1$ (for $d=2,3$) . 
The average residual entanglement of the ancilla qubit $S_\mathrm{anc}$ is finite in the QEC phase and zero in the QZ phase. 
When $t_0=0$ and the full system state $|\Phi'\rangle$ in \eqref{eq:evoANC} is obtained from the state  $|\Psi\rangle\otimes |a\rangle$ where $\Psi$ is the initial product state of the circuit, the scaling of $S_\mathrm{anc}$ determines the boundary order parameter $\beta_s$. In contrast, when $t_0 = 10L$ for $d=2,3$ (whereas, for $d=1$, $t_0 = 4L$) is taken as time after which the state $|\Psi\rangle$ is already in the steady state regime (the average values of observables are time independent), the scaling of $S_\mathrm{anc}$ determines the bulk order parameter critical exponent $\beta$. As anticipated in Sec.~\ref{sec:overview}, in both cases $S_\mathrm{anc}\propto |p-p_c|^{\tilde\beta}$ with $\tilde\beta=\beta,\beta_s$. Calculating $S_\mathrm{anc}$ for several system sizes $L$ for $(d+1)$-dimensional stabilizer circuits, we determine the exponents $\tilde\beta$ to be equal to
\begin{align}
    d=1 & \qquad \beta=0.129(8),\quad \beta_s=0.46(2),\nonumber\\
    d=2 & \qquad \beta=0.44(1),\quad \beta_s=0.86(2),\\
    d=3 & \qquad \beta=0.60(3),\quad \beta_s=1.14(9)\nonumber.
\end{align}
For presentation purposes, we show in Fig.~\ref{fig:3deta} only the results for $\beta_s$, while leaving those for $\beta$ to Appendix~\ref{app:tests}. 
Our estimates for (2+1)D and (3+1)D circuits are compatible within one error bar with those of 3D and 4D percolation (cf.~Table~\ref{tab:critical_exponents}). In particular, we observe that the results for $(2+1)D$ circuits are in contrast with the previous analysis in Ref.~\cite{lunt2021measurementinduced}, where the reported critical exponent $\beta_s$ is incompatible with that of a 3D percolation field theory. We attribute this discrepancy to the limited system sizes in Ref.~\cite{lunt2021measurementinduced}.

\paragraph{Mutual information between ancilla qubits}
Now, we set the measurement rate to its critical value, $p=p_c$ and use the value of the dynamical critical exponent $z=1$, as estimated from the entanglement (Sec.~\ref{sec:results}) and purification observables (Sec.~\ref{sec:purtrans}). This allows us to investigate the bulk-bulk ($\eta$), bulk-boundary ($\eta_\perp$), and boundary-boundary ($\eta_\parallel$) critical exponents~\cite{lunt2021measurementinduced,zabalo2020critical} that describe the behavior of quantum mutual information between two ancilla qubits. To that end, we
entangle the ancilla qubits with the two system qubits, so that at time $t_0$, the state of the full system is given by $|\Phi'\rangle = U_b U_a |\Psi\rangle\otimes |a,b\rangle$, with $U_a$ ($U_b$) entangling the ancilla  qubit$a$ ($b$) and the system qubit at position $\mathbf{r}_a$ ($\mathbf{r}_b$). Subsequently, at times $t>t_0$, we compute the average bipartite quantum mutual information between the two ancilla qubits, namely $I_2 = \mathbb{E}_\mathbf{m}[I_2(\rho_{\mathbf{m},a,b})]$, with $I_2(\rho_{\mathbf{m},a,b})$ given by Eq.~\eqref{eq:i2def}
We can access all the exponents $\tilde\eta\in \{\eta,\eta_\perp,\eta_\parallel\}$ by investigating the dependence of $I_2(r) \sim r^{-(d-1+\tilde\eta)} $ and  varying the boundary conditions and times $t_0$. 
Specifically, we set $t_0=10L$ for $d=2,3$ ($t_0=4L$ for $d=1$) and periodic boundary condition for $\eta$, $t_0=0$ and periodic boundary conditions for $\eta_\parallel$ and $t_0=10L$ for $d=2,3$ ($t_0=4L$ for $d=1$) and open boundary conditions for $\eta_\perp$. The ancilla qubits are initially entangled with system qubits separated by distance $r=L/2$ (for open boundary conditions one of the qubits is at the boundary of the system and one in its geometric center).
Performing simulations for periodic boundary condition and  $t_0=10L$ for $d=2,3$ ($t_0=4L$ for $d=1$), we obtain the quantum mutual information $I_2$ between acilla qubits shown in Fig.~\ref{fig:3deta}(\textbf{d}),(\textbf{e}),(\textbf{f}), respectively for $d=1,2,3$. The value of the bulk-bulk exponent $\eta$ is obtained by optimizing the collapse of  $I_2$ as a function of $(t-t_0)/L$ for different system sizes, see the insets in Fig.~\ref{fig:3deta}(\textbf{d}),(\textbf{e}),(\textbf{f}), we obtain (the analysis for $\eta_\perp$ and $\eta_\parallel$ is presented in  the Appendix~\ref{app:tests})
\begin{align}
    d=1 & \quad \eta=0.212(4),\eta_\perp=-0.02(3),\eta_\parallel=-0.04(5)\nonumber\\
    d=2 & \quad \eta=0.70(2),\eta_\perp=0.99(4),\eta_\parallel=1.5(2)\\
    d=3 & \quad \eta=0.461(8),\eta_\perp=0.43(5),\eta_\parallel=0.42(10)\nonumber .
\end{align}
The obtained values of the critical exponents of $(d+1)$-dimensional circuits are compatible, within the estimated error bars, with the percolation field theory in $(d+1)$ spatial dimensions. 
Again, for the (2+1)D circuits, this contrasts with the conclusion of Ref.~\cite{lunt2021measurementinduced}, possibly due to the smaller system sizes considered in that work. 

\section{Discussion}
\label{sec:dis}
Collecting our numerical findings provides relevant insights into the measurement-induced phase transition in stabilizer circuits. 
Our analysis of the critical exponents, summarized in Table~\ref{tab:critical_exponents} shows the stabilizer circuit measurement-induced phase transition belongs to a universality class close to a percolation field theory. For $d\ge 2$, we find all the critical exponents compatible with those of $(d+1)$ percolation theory.
In particular, the analysis of this manuscript resolves the discrepancies and imprecisions stemming from the limited system sizes in previous studies of the (2+1)D systems. In contrast with Ref.~\cite{turkeshi2020measurementinduced}, we show that (2+1)D stabilizer circuits have a critical point with area-law scaling of the entanglement entropy and scaling invariance ($z=1$, cf. Eq.~\eqref{eq:resz}). Moreover, in contrast to ~\cite{turkeshi2020measurementinduced}, we find that the critical exponent $\nu$ for (2+1)D circuits is compatible with the critical exponent for 3D classical percolation, confirming the result of \cite{lunt2021measurementinduced}. In contrast with Ref.~\cite{lunt2021measurementinduced}, we find the surface critical exponents compatible with those of 3D percolation theory. 
Within one error bar, our numerics verify the compatibility of the critical exponents between two and three-dimensional stabilizer circuits with those of 3D and 4D percolation theory. 

\begin{figure}[t!]
    \centering
    \includegraphics[width=0.9\columnwidth]{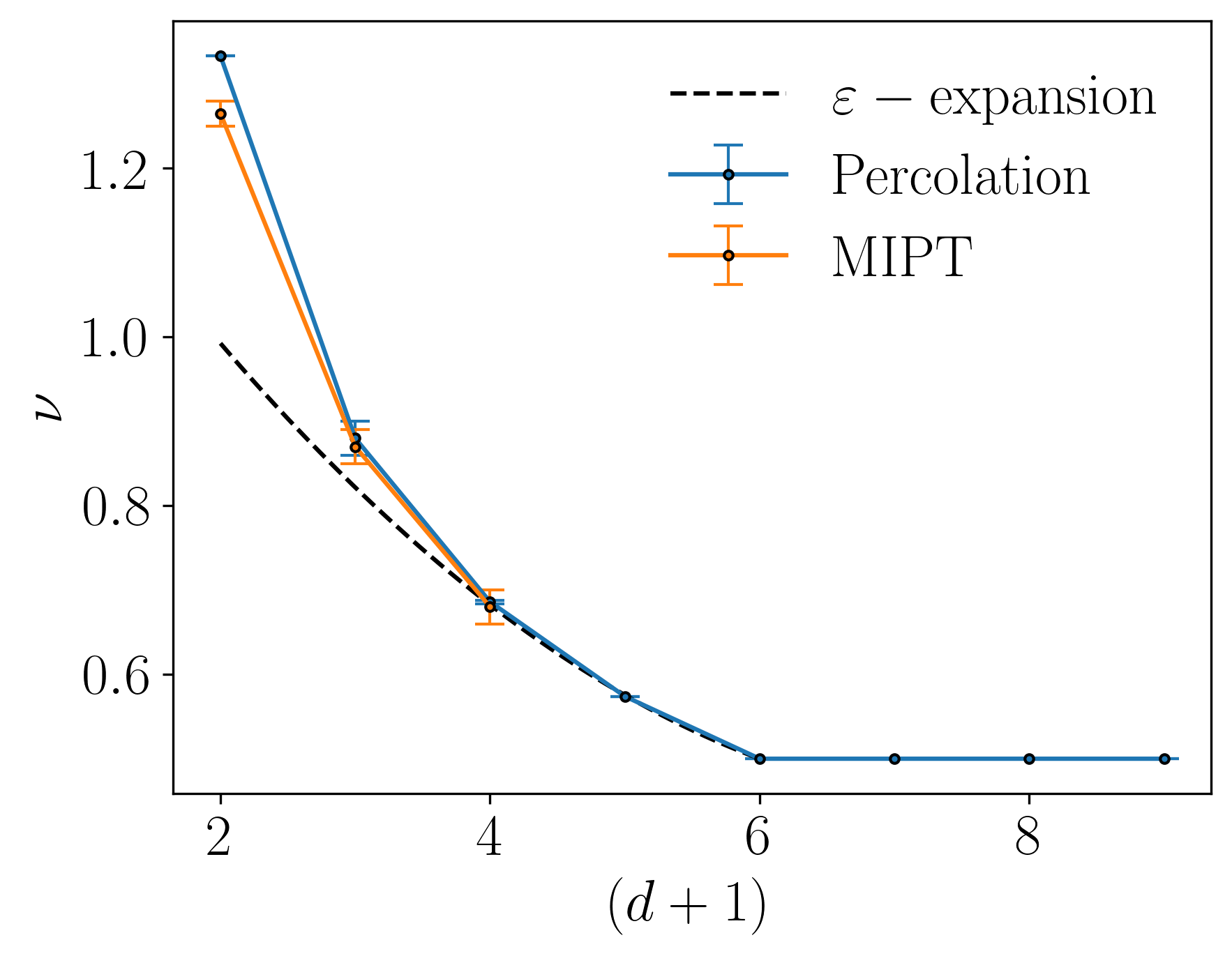}
    \caption{Correlation length critical exponent $\nu$ for various dimensions $d$. As argued throughout this paper, the critical exponents of measurement-induced phase transition in $(d+1)$ space-time dimensions are close to the analogous exponents for classical percolation in $(d+1)$ spatial dimensions. As a reference, we consider the values of $\nu$ in Ref.~\cite{phifiveloops}. The black dashed line is the result of the renormalization group $\varepsilon$ expansion~\cite{phifiveloops}, which we use as a guiding eye for $d\ge 3$. A similar behavior arises for the other critical exponents (cf. Table~\ref{tab:critical_exponents}). }
    \label{fig:comparisonRG}
\end{figure}
To illustrate this point, we report in Fig.~\ref{fig:comparisonRG} a comparison between the correlation length critical exponent in stabilizer circuits in $(d+1)$ spacetime dimensions and in percolation field theory in $(d+1)$ space dimensions for various values of $d$. 
Here, we also present the perturbative estimate of $\nu$ renormalization group $\varepsilon$-expansion at $d=6-\varepsilon$ for the percolation field theory~\cite{phifiveloops}
\begin{equation}
    \nu^\mathrm{RG}(d=6-\varepsilon) = \frac{1}{2} + \frac{5}{84}\varepsilon+\frac{589}{37044}\varepsilon^2+O(\varepsilon^3).
\end{equation}
Due to the agreement for $d\ge 2$, we propose a conjecture that the universality class of MIPT in $(d+1)$D circuits lies in the vicinity of $(d+1)$ percolation theory even in higher dimensions, $d\ge 4$, not considered in this paper. In particular, it is evocative to conjecture the upper critical dimension of percolation field theory, $d_c=6$ matches that of stabilizer circuits.

To contextualize these results, we invoke the statistical mechanics mapping in Ref.~\cite{li2021statistical_1}. There, the authors consider the $q$-qudit generalization of the (1+1)D stabilizer circuit in Sec.~\ref{sec:methods} and map it to a statistical mechanics model on a honeycomb lattice
\begin{equation}
    Z = \sum_{\{s\in \Sigma\}} e^{-\beta \mathcal{H}(\{s\}) },\label{eq:modelss}
\end{equation}
where $s$ are effective variables defined on a set $\Sigma$ and $\mathcal{H}(\{s\})$ are Boltzmann weights. Eq.~\eqref{eq:modelss} is in general hard to treat since the weights are not positively defined. 
Nevertheless, the authors show the model maps exactly to percolation field theory on a square lattice in the limit of large local Hilbert space dimension $q\to\infty$, with 
\begin{equation}
    \mathcal{H} = \sum_{\langle i,j\rangle } \delta_{s_i,s_j},
\end{equation}
where the sum is over neighboring sites. Instead, for large but finite on-site Hilbert space dimension, $\infty >q\gg1 $, the system is perturbed by a relevant operator, that possibly moves the universality class away from the percolation field theory.
A similar argument can be extended to higher dimensions, $d\geq2$. Since the weights in \eqref{eq:modelss} are fixed by the measurement operations and by the unitary gates, we expect the large qudit limit of $(d+1)$-dimensional stabilizer circuits will be again described by a perturbed percolation theory. The main difference is the lattice for $d\ge 2$ has a more involved structure and is different for every model considered in Sec.~\ref{sec:methods} and Appendix~\ref{app:tests}. 
Two remarks are in order here. First, from the representation theory~\cite{Vasseur2014}, these relevant perturbations are expected to have the same scaling dimensions of the energy operator, hence they should be relevant for any $d$.  Notably, this argument holds also beyond the mean-field theory $d\ge d_c$.
Furthermore, for stabilizer circuits built on qubits ($q=2$), the effects of this relevant operator are expected to be the strongest. 

Surprisingly, our numerical findings reveal that these effects are very mild in $d\ge 2$, and as previously stressed we cannot differentiate between the critical exponents of measurement-induced phase transition in the stabilizer circuits and the critical exponents of the percolation field theory.
We leave this open dilemma for future investigations.

We conclude by discussing the conformal field theory ansatz of the critical point. This assumption guided the choice of scaling function in Eq.~\eqref{eq:pur}, and here we present evidence the measurement-induced critical points of stabilizer circuits are indeed conformal. 
The emergence of conformal invariance at (1+1)D MIPT has been already thoroughly studied in \cite{li2021conformal}. In particular, the bipartite quantum mutual information between two subsystems of (1+1)D circuit has been shown to be solely a function of cross ratios dictated by CFT \cite{li2019measurementdriven}. However, the situation for $d\geq 2$ is more involved. 
To approach this problem, we investigate the torus entanglement entropy, i.e. the average entanglement entropy of the region $A$ assuming periodic boundary conditions in the system in Fig.~\ref{fig:partition}. Assuming that conformal field theory holds, the torus entanglement entropy is given by \cite{Witczak17}
\begin{equation}
    S_A= \mathcal{B} L_y - \chi(b, L_A/L_x) + \mathcal O(1/L_y),
    \label{eq:s2}
\end{equation}
for (2+1)D spatial dimensions and 
\begin{equation}
    S_A= \mathcal{B} L_yL_z - \chi(b, L_A/L_x) + \mathcal O(1/L_y),
    \label{eq:s3}
\end{equation}
for (3+1)-dimensional system, where $b=L_x/L_y$, $\mathcal{B}$ is a prefactor that characterizes the area-law scaling of $S_A$, whereas the term $\chi$ depends non-trivially only on the aspect ratios characterizing the geometry of the problem (we keep $L_y=L_z$ for $d=3$).

\begin{figure}[t!]
    \centering
    \includegraphics[width=0.9\columnwidth]{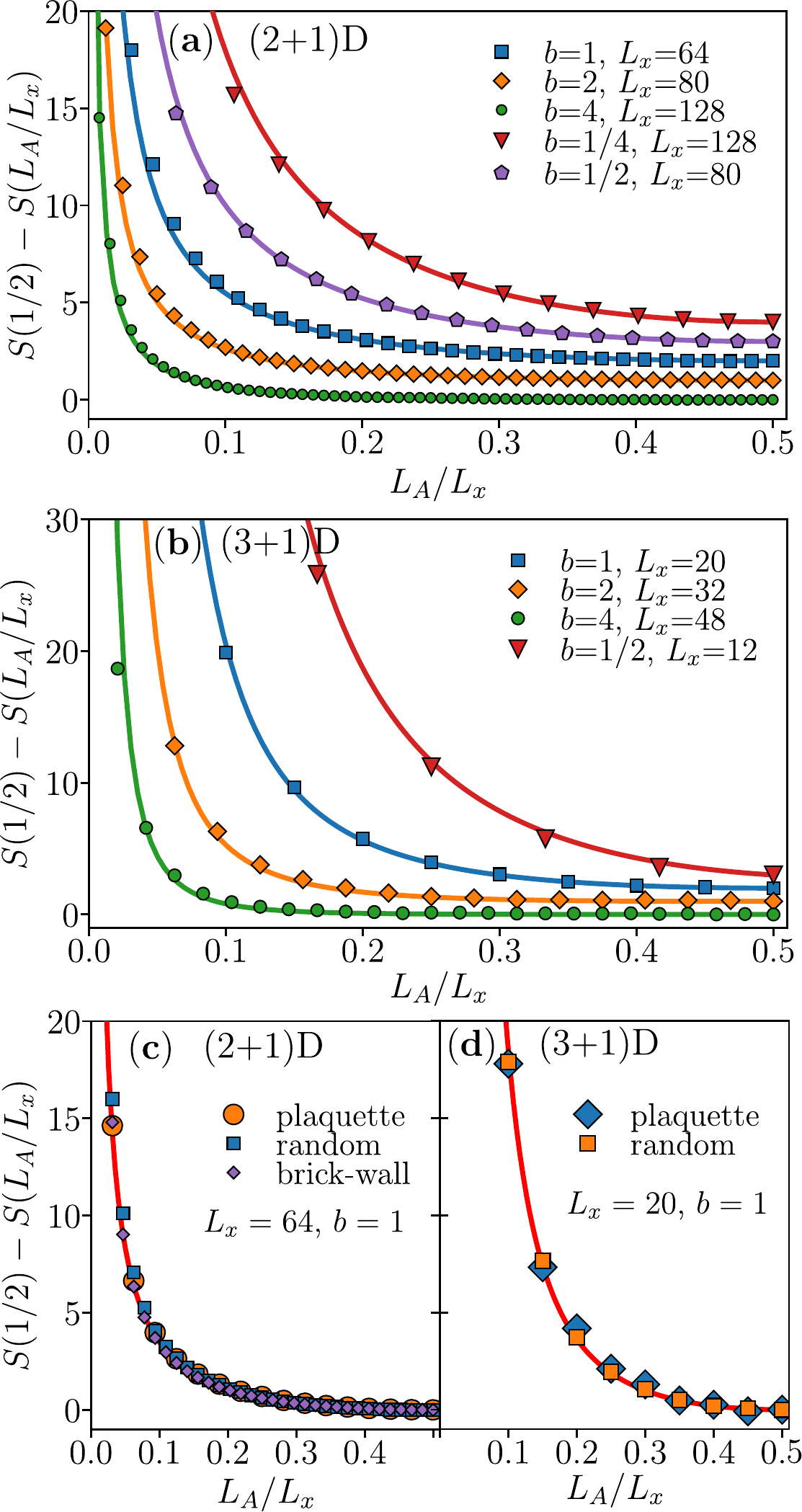}
    \caption{Dependence of entanglement entropy at MIPT ($p=p_c$) on subsystem size $L_A$  for (2+1)D (\textbf{a}) and (3+1)D (\textbf{b}) circuits of fixed aspect ratios $b = L_x/L_y$ ($L_y=L_z$ for (3+1)D case). Results for random circuits are compared with predictions for systems with conformal symmetry in 2 and 3 spatial dimensions derived from the EMI model (denoted by solid lines). The panel (\textbf{c}) shows a comparison between (2+1)D circuits with different spatiotemporal structures:  data for plaquette and brick-wall circuits are rescaled by factors $0.68$ and $0.88$, respectively. Analogous comparison for (3+1)D circuits is shown in the panel (\textbf{d}), and results for the plaquette circuit are rescaled by a factor $0.60$.
   } 
    \label{fig:torusENT}
\end{figure}

We numerically compute the entanglement entropy of subsystem $A$, $S_A \equiv S(L_A/L_x)$ varying $b$, $L_x$, and $L_A$ for (2+1)D and (3+1)D stabilizer circuits. To isolate the sub-leading term $\chi(b, L_A/L_x)$, we consider the difference $S(1/2)-S(L_A/L_x)$. We observe that $S(1/2)-S(L_A/L_x)$ plotted as function of $L_A/L_x$ for several values of $b$ follow the same curves for two distinct values of $L_x$, verifying the CFT scalings \eqref{eq:s2}, \eqref{eq:s3} for both (2+1)D and (3+1)D circuits. Moreover, we take the larger value of $L_x$ and compare the obtained values of $S(1/2)-S(L_A/L_x)$ with the same quantity calculated for conformally invariant Extensive Mutual Information (EMI) model~\cite{Casini05,Casini09} (the analytical expressions are given in \cite{Witczak17}). We find an accurate agreement between the results obtained for (2+1)D and (3+1)D stabilizer circuits at the critical point, $p=p_c$, and the results for $S(1/2)-S(L_A/L_x)$ in the EMI model, as shown in Fig.~\ref{fig:torusENT}(\textbf{a}) and (\textbf{b}). We note that the analytical expressions for the EMI model contain a single proportionality factor $\kappa$ which we use as a fitting parameter. The value of $\kappa$ depends on the dimensionality of the system and the circuit architecture, cf. Fig.~\ref{fig:torusENT}(\textbf{c}) and (\textbf{d}), but is independent of the parameters characterizing the geometry of the system. Finally, with increasing $L_x$, we observe an increase of the width of the interval of $L_A/L_x$ for which the agreement between $S(1/2)-S(L_A/L_x)$ for stabilizer circuits at the critical point and EMI model is observed. 

This analysis provides suggestive evidence that the measurement-induced transition for $(d+1)$-dimensional stabilizer circuits is described by an emergent conformal field theory.

\section{Conclusion}
\label{sec:conclusion}
This paper investigated measurement-induced phase transitions in stabilizer circuits in $(d+1)$ space-time dimensions. Using large-scale numerical simulations, we demonstrated that the universality class of measurement-induced phase transition is close to that of percolation field theory in $(d+1)$ spatial dimensions. 

For (1+1)D systems, the distinction between critical exponents of MIPT and 2D percolation conformal field theory is achieved thanks to larger system sizes with respect to previous studies in the literature (e.g. see Ref.~\cite{zabalo2020critical}). 
In higher dimensions, our numerics evidence that the critical point of the measurement-induced phase transition is still captured by a conformal theory and that the critical exponents are compatible within one error bar with those of percolation field theory. 
In particular, this work resolved the imprecise statements in Ref.~\cite{turkeshi2020measurementinduced,lunt2021measurementinduced} for (2+1)D systems, while providing the first characterization of MIPT in (3+1)D stabilizer circuits. We have emphasized the complementarity of the alternative approaches to the measurement-induced criticality based on the entanglement measures, on the purification of initial mixed states, and on the participation entropy of wave functions on individual quantum trajectories. Importantly, we have confirmed that this equivalence extends to dimensions beyond $d=1$.

Our work leaves some open questions. In view of the analytical considerations, it is not clear what is the mechanism behind the closeness (or coincidence) of percolation field theory and MIPTs in $d\ge 2$. In spite of analytical arguments based on large local Hilbert space dimensions suggesting these universality classes should be distinct, our numerics show their compatibility even for qubits, for which these differences should be the most pronounced. We believe that more refined analytical methods are in order to shed light on this issue.

There are several interesting future directions. First, it would be interesting to study the stabilizer circuits with all-to-all and quantum trees interactions~\cite{nahum2020measurement}, and in complex geometry stabilizer tensor networks~\cite{li2021statistical_1}. 
Furthermore, large-scale numerical simulations can help understand the nature of the error-correcting phase in higher dimensions. For example, in (1+1)-dimensional circuits, it is known that stabilizer circuits present Kardar-Parisi-Zhang fluctuations with the typical $1/3$-exponent~\cite{weinstein,li2021statistical_2}. It would be interesting to extend the statistical field theory of stabilizer circuits beyond one spatial dimension and to characterize the quantum fluctuations in the quantum error-correcting phase. 
Additionally, it would be interesting to study three-dimensional measurement-protected orders, such as the topological phase transitions in stabilizer circuits~\cite{lavasani2021topological,sang2021measurementprotected}.

Lastly, it would be interesting to devise schemes to detect MIPTs for $d>1$. The main obstacle is the fact that the probability of ending up with a certain output state of the circuit \eqref{eq:deftraj} is exponentially small in the number of performed measurements, yielding exponentially large system size number of circuit realizations required to repeatedly probe the same output state. This is the so-called problem of post-selection and is particularly severe for $d>1$. Some recent works in proposed various approaches to circumvent this problem~\cite{Iadecola22, Buchhold22, Friedman22}. It would be interesting to see whether those schemes can be adapted to higher dimensions and may overcome the brute force realization in Ref.~\cite{Noel_2022,ibmmotta}, where the post-selection was performed. 

\begin{acknowledgments}
PS and XT thank R. Vasseur for enlightening discussions.
XT and MS acknowledge support from the ANR grant “NonEQuMat”
(ANR-19-CE47-0001) and computational resources on the Coll\'ge de France IPH cluster. 
ICFO group acknowledges support from: ERC AdG NOQIA; Ministerio de Ciencia y Innovation Agencia Estatal de Investigaciones (PGC2018-097027-B-I00/10.13039/501100011033, CEX2019-000910-S/10.13039/501100011033, Plan National FIDEUA PID2019-106901GB-I00, FPI, QUANTERA MAQS PCI2019-111828-2, QUANTERA DYNAMITE PCI2022-132919, Proyectos de I+D+I “Retos Colaboración” QUSPIN RTC2019-007196-7); MCIN Recovery, Transformation and Resilience Plan with funding from European Union NextGenerationEU (PRTR C17.I1); Fundació Cellex; Fundació Mir-Puig; Generalitat de Catalunya (European Social Fund FEDER and CERCA program (AGAUR Grant No. 2017 SGR 134, QuantumCAT \ U16-011424, co-funded by ERDF Operational Program of Catalonia 2014-2020); EU Horizon 2020 FET-OPEN OPTOlogic (Grant No 899794); National Science Centre, Poland (Symfonia Grant No. 2016/20/W/ST4/00314); European Union’s Horizon 2020 research and innovation programme under the Marie-Skłodowska-Curie grant agreement No 101029393 (STREDCH) and No 847648 (“La Caixa” Junior Leaders fellowships ID100010434: LCF/BQ/PI19/11690013, LCF/BQ/PI20/11760031, LCF/BQ/PR20/11770012, LCF/BQ/PR21/11840013).
The authors thankfully acknowledge the computer resources at MareNostrum and the technical support provided by Barcelona Supercomputing Center (FI-2022-1-0042).
\end{acknowledgments}

\appendix 

\begin{figure*}[t!]
    \centering
    \includegraphics[width=\linewidth]{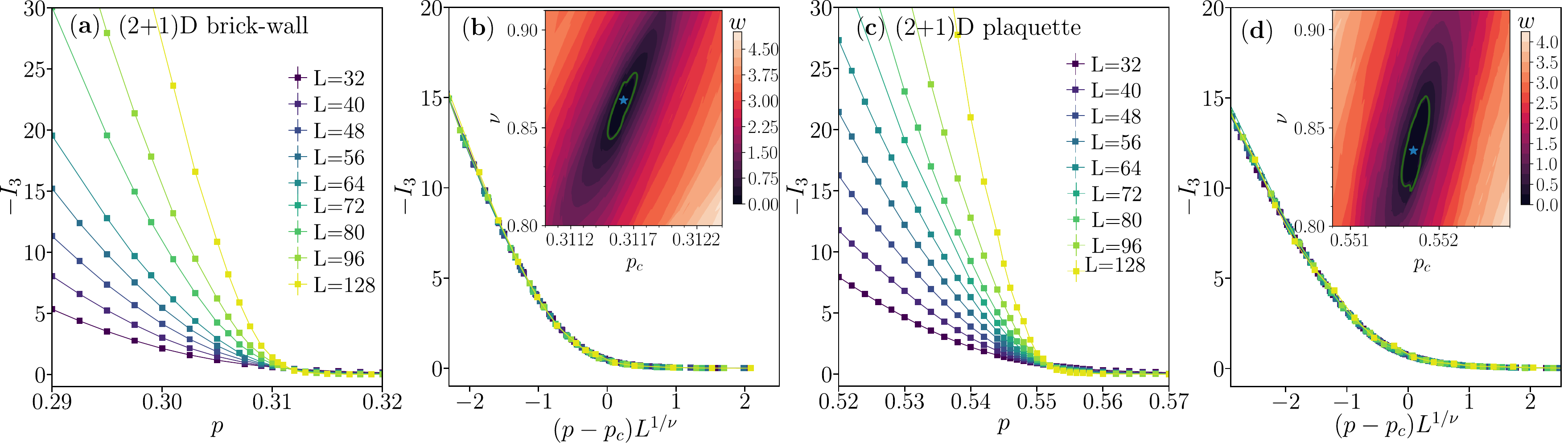}
    \caption{Additional data on entanglement transition in (2+1)D circuits. (\textbf{a}) TQMI $I_3$ as a function of  measurement rates $p$ for various system sizes $L$ for the brick-wall circuit architecture. (\textbf{b}) Data collapse of  $I_{3}$ for the critical parameters $p_c=0.3116(1)$, $\nu=0.863(16)$, the inset shows landscape of the cost function $w$ used to locate the critical parameters. (\textbf{c}) TQMI $I_{3}$ as a function of $p$ for various $L$ for the plaquette circuit architecture. \textbf{d}) Data collapse of  $I_{3}$ for the critical parameters $p_c=0.5517(1)$, $\nu= 0.839(22)$, the inset shows landscape of the cost function $w$.}
    \label{fig:2d_additional}
\end{figure*}

\begin{figure}[t!]
    \centering
    \includegraphics[width=0.85\linewidth]{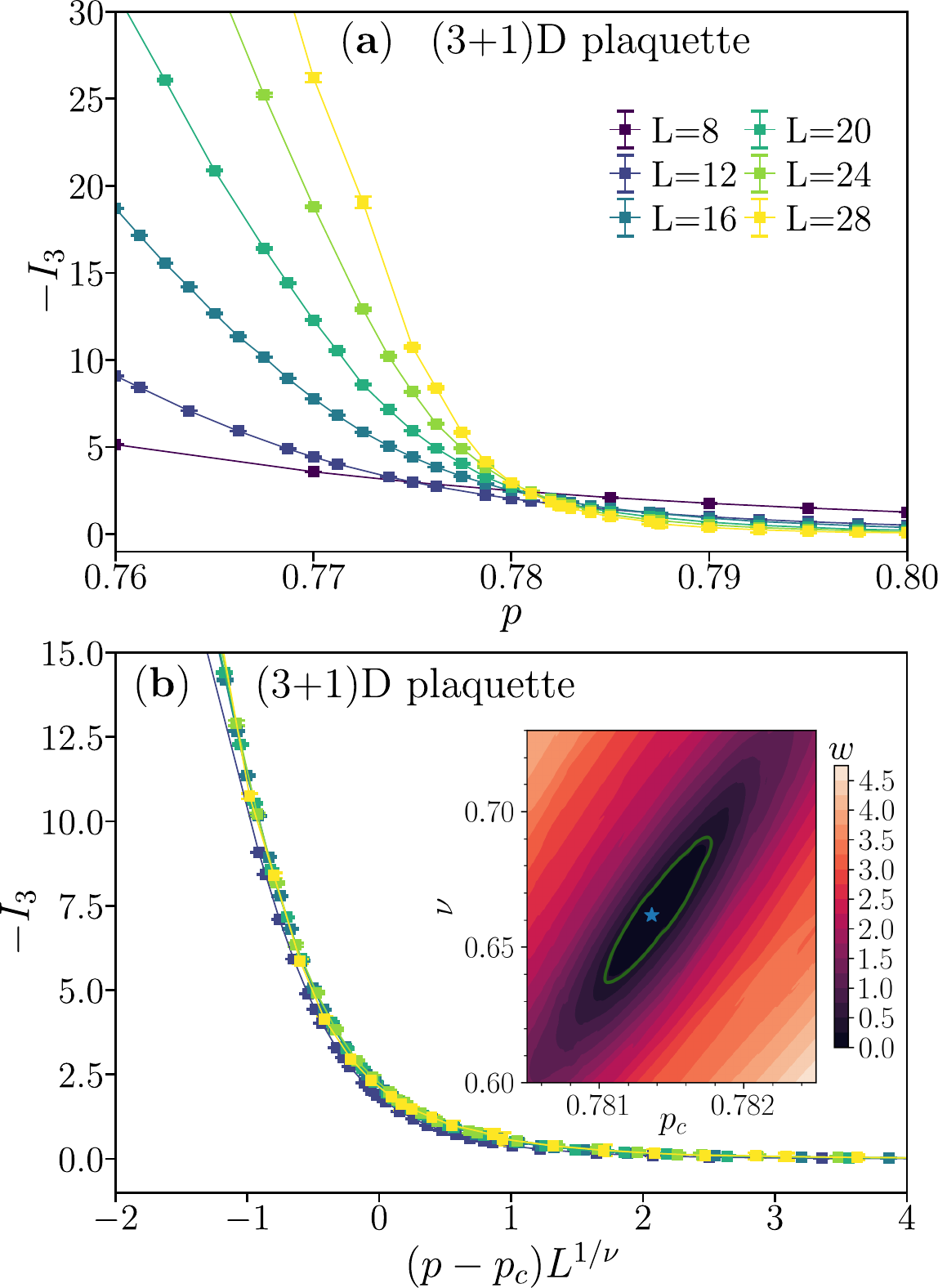}
    \caption{Additional data on entanglement transition in (3+1)D circuits. (\textbf{a}) TQMI $I_3$ as a function of  measurement rates $p$ for various system sizes $L$ for the plaquette circuit architecture. (\textbf{b}) Data collapse of  $I_{3}$ for the critical parameters $p_c=0.7814(4)$, $\nu= 0.662(22)$, the inset shows landscape of the cost function $w$ used to locate the critical parameters.}
    \label{fig:3d_additional}
\end{figure}

\section{Benchmark of universal behavior}
\label{app:tests}
In this Appendix, we discuss the universality of our numerical results for the critical exponents employing various circuital architectures. 
Specifically, we discuss: (i) the data collapse for the tripartite mutual information for the (2+1)D circuital architectures discussed in Ref.~\cite{turkeshi2020measurementinduced,lunt2021measurementinduced} and the data collapse for the tripartite mutual information for a (3+1)D regular circuit, and (ii) the data collapse for $\beta,\eta_\parallel,\eta_\perp$.

\subsection{Tests of universal behavior for (2+1)D and (3+1)D circuits}
We perform the numerical simulations within the architectures in Ref.~\cite{turkeshi2020measurementinduced,lunt2021measurementinduced} to prove our results are independent of the specific microphysics.

In Ref.~\cite{lunt2021measurementinduced} the authors study a circuit architecture with the measurement layer in Eq.~\eqref{eq:measlayer} and the brick-wall unitary layer built upon two-body random Clifford gates.
Two indices determine the pattern of the gates: a sublattice index $\mathcal{A}=\textrm{mod}(i_x+i_y,2)$ (where $\vec{i}=(i_x,i_y)$) that determine which sublattice of qubits will act as the controls for the Clifford gates and a clock index $\mathcal{Q}\in \{0,1,2,3\}$ that fix the direction of the two-body gates. The values $q=0,1,2,3$ correspond to adding the versor $\hat{r}(q) =\hat{e}_y,\hat{e}_x,-\hat{e}_y,-\hat{e}_x$, respectively. We denote $\Lambda_{a}$ the sublattice fixed by $\mathcal{A}=a$.
The time evolution is given by 
\begin{equation}
    U(t) = \prod_{\vec{i}\in \Lambda_{\textrm{mod}(t,2)}} U(\vec{i}+\hat{r}(\textrm{mod}(\lfloor t/2\rfloor,4)),t),
\end{equation}
where $\lfloor n\rfloor $ is the integer part of $n$.
We simulate the circuit up to $t=4L$ and study the behavior of $I_3$. Our The results are given in Fig.~\ref{fig:2d_additional}(a,b). The critical point and correlation length exponent are $p_c=0.3116(1)$ and $\nu=0.863(16)$ and are compatible with those in Ref.~\cite{lunt2021measurementinduced} and with Eq.~\eqref{eq:res2drand}.

Instead, in Ref.~\cite{turkeshi2020measurementinduced} the authors consider alternating measurement and unitary layers, with measurements given as in Eq.~\eqref{eq:measlayer}, while the unitary is fixed by the plaquette layer built upon the 4-body random Clifford gates
\begin{equation}
    V^{(4)}(i_x,i_y,t) \equiv U_{\vec{i},\vec{i}+\hat{e}_x,\vec{i}+\hat{e}_y,\vec{i}+\hat{e}_x+\hat{e}_y}
\end{equation}
where $\vec{i}=(i_x,i_y)\in\Lambda$ and $\hat{e}_\alpha$ is the unit versors along the direction $\alpha=x,y$.
There are specified by 
\begin{equation}
\begin{split}
    U(t) &= \prod_{i_x=1}^{L_x/2}\prod_{i_y=1}^{L_y/2} V^{(4)}(2i_x-r_x(t),2i_y-r_y(t),t)\\
    r_x(t)&=\begin{cases}
        1 & \text{if } \mod(t,4)=1,2,\\
        0 & \text{otherwise},\\
    \end{cases}\\
    r_y(t)&=\begin{cases}
        1 & \text{if } \mod(t,4)=0,1,\\
        0 & \text{otherwise}.\\
    \end{cases}
\end{split}
\label{eq:ziocan}
\end{equation}
We let the system evolve up to time $t=4L$ and compute the tripartite mutual information (cf. Sec.~\ref{sec:results}). The results are given in Fig.~\ref{fig:2d_additional}(c,d). We see the data collapse is given by $p_c=0.5517(1)$ and $\nu=839(22)$. In particular, the correlation length critical exponent is compatible with Eq.~\eqref{eq:res2drand}.

\begin{figure*}[t!]
    \centering
    \includegraphics[width=\linewidth]{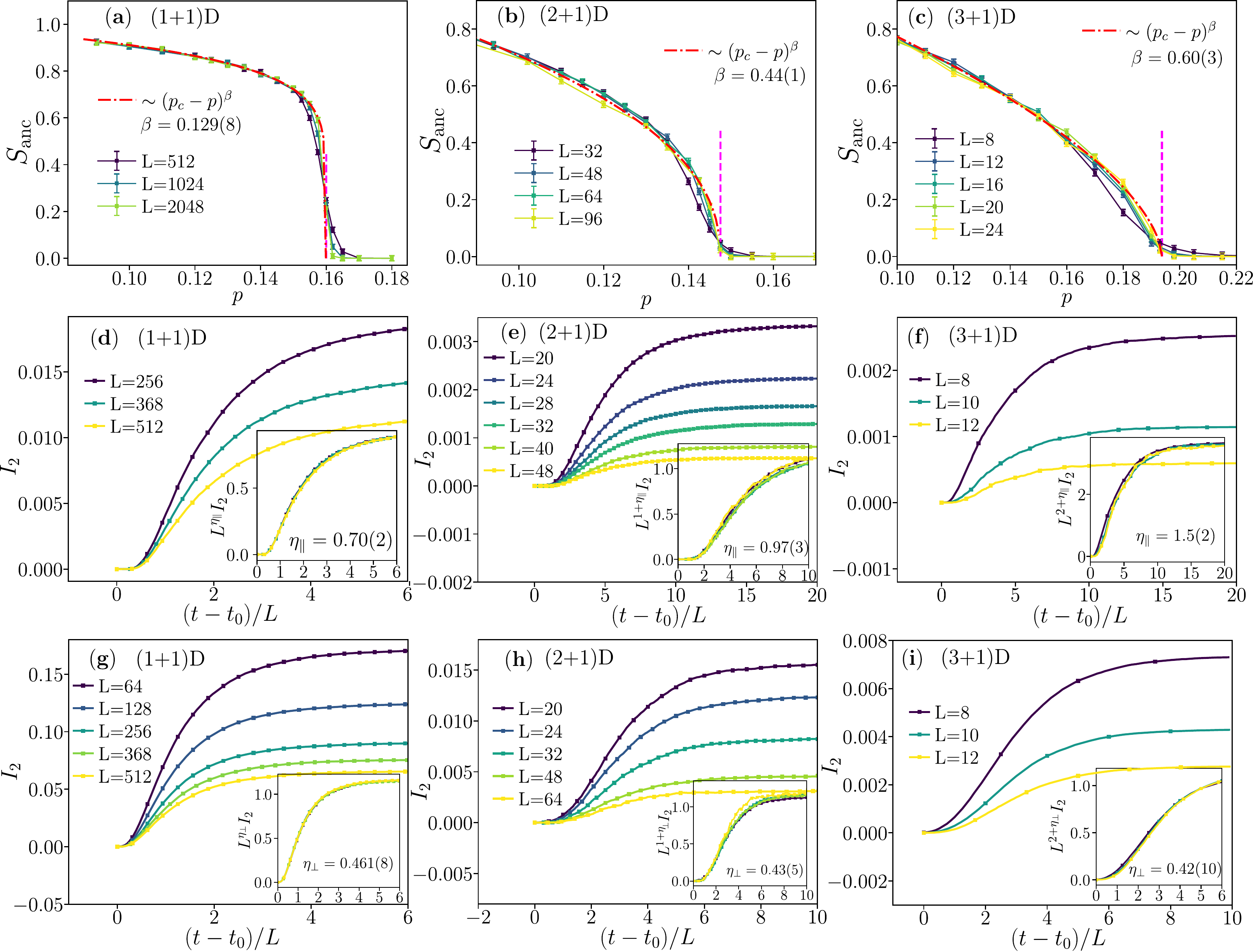}
    \caption{Additional data on the bulk and boundary properties of MIPT. The first row, (\textbf{a})--(\textbf{c}) Calculation of the exponent $\beta$ for (1+1)D, (2+1)D and (3+1)D circuits. The second row, (\textbf{d})--(\textbf{f}) Calculation of the exponent $\eta_{\parallel}$. The third row, (\textbf{g})--(\textbf{i}) Calculation of the exponent $\eta_{\perp}$. }
    \label{fig:bulk_surf_add}
\end{figure*}

Lastly, we consider a different (3+1)D stabilizer circuit, with measurement layers as in Eq.~\eqref{eq:measlayer} and unitary layers built upon the 8-body random Clifford gates
\begin{equation}
    V^{(8)}(i_x,i_y,i_z,t) \equiv U_{\vec{i},\vec{i}+\hat{e}_x,\vec{i}+\hat{e}_y,\vec{i}+\hat{e}_x+\hat{e}_y}^{ \vec{i}+\hat{e}_z,\vec{i}+\hat{e}_x+\hat{e}_z,\vec{i}+\hat{e}_y+\hat{e}_z,\vec{i}+\hat{e}_x+\hat{e}_y+\hat{e}_z}
\end{equation}
where $\vec{i}=(i_x,i_y,i_z)\in\Lambda$ and $\hat{e}_\alpha$ is the unit versors along the direction $\alpha=x,y,z$.
The unitary layer is then specified by
\begin{equation}
    U(t) = \prod_{i_x=1}^{L_x/2}\prod_{i_y=1}^{L_y/2} \prod_{i_z=1}^{L_z/2}V^{(8)}(2i_x-r_x(t),2i_y-r_y(t),2i_z-r_z(t),t)
\end{equation}
with $r_x(t)$ and $r_y(t)$ as in Ref.~\eqref{eq:ziocan}, while $r_z(t) = \lfloor\mathrm{mod}(t,8)/4\rfloor$. 

We let the system evolve up to time $t=L$ and compute the TQMI. The results are given in Fig.~\ref{fig:3d_additional}. The optimal parameters are $p_c=0.7814(4)$ and $\nu=0.662(22)$. As for the two-dimensional circuits, these correlation length critical exponent is compatible with the results on randomized circuits (cf. Eq.~\eqref{eq:3dest}). 

\subsection{Bulk and boundary critical exponents}
As described in Sec.~\ref{sec:blkbnd}, here we present the data collapses for the $(d+1)$-dimensional stabilizer circuits. All our estimates agree with the ones of percolation theory in $(d+1)$-spatial dimensions.

\onecolumngrid
\section{Details of the finite-size scaling procedure}
\label{app:ffs}
In this Appendix, we detail how we perform the finite size scaling to obtain the critical point and the critical exponents~\cite{Kawashima1993}. In general, we organize the data in pairs $(x_i,y_i,d_i)$, where $i=(p,L,t)$ denote the raw data indices for the scaling observables $x_i$, the conditional averaged observable $y_i$ and the errors $d_i=\sigma_{y_i}$. 
We note that both the $x_i(\boldsymbol{\alpha})$ and the $y_i(\boldsymbol{\gamma})$ may depend on the critical point and exponents $\{\alpha_i\}$, $\{\gamma_i\}$. 
(For instance, $x_i=(p-p_c)L^{1/\nu}$ at equal time $t$, and $y_i=I_3(p,L,t)$.)

The cost function determines the quality of the collapse
\begin{equation}
\label{eq:cost}
    \mathcal{W}(p_c,\boldsymbol\alpha,\boldsymbol\gamma)\equiv \frac{1}{n-2}\sum_{i=2}^{n-1} \mathcal{M}(x_i,y_i,d_i|x_{i-1},y_{i-1},d_{i-1},x_{i+1},y_{i+1},d_{i+1}),
\end{equation}
where $n$ is the total number of data, and we have sorted the data such that $x_1<x_2<\dots<x_n$, and the local weight $\mathcal{M}(x,y,d|x',y',d',x'',y'',d'')$ is given by
\begin{align}
    \mathcal{M} &= \left(\frac{y-\overline{y}}{\Delta(y-\overline{y})}\right)^2, \quad
    \overline{y} = \frac{(x''-x)y' - (x'-x)y''}{x''-x'},\quad 
    |\Delta(y-\overline{y})|^2  = d^2 + \left(\frac{x''-x}{x''-x'}d'\right)^2 + \left(\frac{x'-x}{x''-x'}d''\right)^2.
\end{align}
Heuristically, $\mathcal{W}$ minimizes the difference of the data curves for various $p$, $L$, and $t$, with unity as the ideal minimum value. We evaluate $\mathcal{W}$ for different choices of parameters, and identify the optimal collapse as the minimum $(\boldsymbol{\alpha}^\star,\boldsymbol{\gamma}^\star)$. The error is identified as the region for which $\mathcal{W}(\boldsymbol{\alpha},\boldsymbol{\gamma})\le 1.3 \mathcal{W}(\boldsymbol{\alpha}^\star,\boldsymbol{\gamma}^\star)$~\cite{zabalo2020critical}. 
\twocolumngrid
%

\bibliographystyle{apsrev4-2}

\end{document}